\documentclass[letter,longauth]{aa} 
      
\usepackage{upgreek} 
\usepackage{color}
\usepackage{hhline}
\usepackage{longtable}
\usepackage{graphicx}

\usepackage[varg]{txfonts}
\usepackage{upgreek} 
\usepackage{multirow} 
\usepackage[FIGTOPCAP, center, nooneline]{subfigure}
\usepackage{natbib}

\usepackage[varg]{txfonts}
\usepackage{microtype}
\usepackage{hyperref}
\usepackage{booktabs}
\usepackage{xcolor}
\usepackage{float}

\hypersetup{%
  colorlinks=true,
  linkcolor=blue,
  citecolor=blue
}

\usepackage{etoolbox}
\makeatletter

\let\oldcitet=\citet

\renewcommand{\citet}[1]{\textcolor[rgb]{0,0,1}{\oldcitet{#1}}}

\newcommand{\OI}{O\,{\sc i}}

\newcommand{\CI}{C\,{\sc i}}
\newcommand{\SII}{S\,{\sc ii}}

\newcommand{\FeII}{Fe\,{\sc ii}}

\newcommand{\HII}{H\,{\sc ii}}
\newcommand{\HI}{H\,{\sc i}}
\newcommand{\HeI}{He\,{\sc i}}

\begin{document} 


\title{PDRs4All. X. ALMA and JWST detection of neutral carbon in the externally irradiated disk d203-506: Undepleted gas-phase  carbon}

\titlerunning{Neutral carbon in the externally photoevaporating disk d203-506} 
\authorrunning{Goicoechea et al.}

 \author{Javier R.\,Goicoechea\inst{1}
          \and
         \mbox{J. Le Bourlot\inst{2}}
          \and
         \mbox{J. H. Black\inst{3}}
         \and 
         \mbox{F. Alarc\'on\inst{4}}
         \and
         \mbox{E. A. Bergin\inst{4}}
         \and
         \mbox{O. Bern\'e\inst{5}}
         \and
         \mbox{E. Bron\inst{6}}
         \and
         \mbox{A. Canin\inst{5}}
         \and
         \mbox{E. Chapillon\inst{7}}
         \and
         \mbox{R. Chown\inst{8,9,10}}
         \and
         \mbox{E. Dartois\inst{11}}
         \and
         \mbox{M. Gerin\inst{6}}
         \and
         \mbox{E. Habart\inst{12}}
         \and
         \mbox{T. J. Haworth\inst{13}}
         \and
         \mbox{C. Joblin\inst{5}}
         \and
         \mbox{O. Kannavou\inst{12}}
         \and
         \mbox{F. Le Petit\inst{6}}
         \and
         \mbox{T. Onaka\inst{14}}
         \and 
         \mbox{E. Peeters\inst{8,9,15}}
         \and
         \mbox{J. Pety\inst{6,7}}
         \and
         \mbox{E. Roueff\inst{6}}
         \and
         \mbox{A. Sidhu\inst{8,9}}
         \and
         \mbox{I. Schroetter\inst{5}}
         \and
         \mbox{B. Tabone\inst{12}}
         \and
         \mbox{A. G. G. M. Tielens\inst{16,17}}
         \and
         \mbox{B. Trahin\inst{12}}
         \and
         \mbox{D. Van De Putte\inst{18,8,9}}
         \and
         \mbox{S. Vicente\inst{19}}
         \and
         \mbox{M. Zannese\inst{12}}
      }

\institute{Instituto de F\'{\i}sica Fundamental
     (CSIC). Calle Serrano 121-123, 28006, Madrid, Spain.\email{javier.r.goicoechea@csic.es}
\and
Universit\'e Paris Cit\'e, France.
\and
Chalmers University of Technology, Onsala Space Observatory, \mbox{Onsala}, Sweden.
\and
Dept. of Astronomy, University of Michigan, Ann Arbor, MI, USA.
\and
Institut de Recherche en Astrophysique et Planétologie, Universit\'e Toulouse III - Paul Sabatier, CNRS, CNES, Toulouse, France.
\and
LERMA, Observatoire de Paris, PSL Research University, CNRS, Sorbonne Universit\'es, Paris-Meudon, France.
\and
IRAM, 300 rue de la Piscine, 38406 Saint Martin d’H\`eres, France.
\and
Department of Physics and Astronomy, University of Western Ontario, London, Ontario, Canada.
\and
Institute for Earth and Space Exploration, University of Western Ontario, London, Ontario, Canada.
\and
Department of Astronomy, The Ohio State University, 140 West 18th Avenue, Columbus, OH 43210, USA.
\and
Institut des Sciences Mol\'eculaires d’Orsay, CNRS, Universit\'e Paris-Saclay, Orsay, France.
\and
Universit\'e Paris-Saclay, CNRS, Institut d’Astrophysique Spatiale, Orsay, France.
\and
School of Physics and Astronomy, Queen Mary University of London, London E1 4NS, UK. 
\and
Department of Astronomy, Graduate School of Science, University of Tokyo, Tokyo, Japan.
\and
Carl Sagan Center, SETI Institute, Mountain View, CA, USA.
\and
Leiden Observatory, Leiden University, The Netherlands.
\and
Astronomy Department, Univ. of Maryland, College Park, MD, USA.
\and
Space Telescope Science Institute, Baltimore, MD, USA.
\and
Instituto de Astrof\'{\i}sica e Ci\^encias do Espaço, Lisbon, Portugal.
}

   \date{Received 4 June 2024 / Accepted 7 August 2024}

\abstract{The gas-phase  abundance of carbon, 
$x_{\rm C}$\,=\,[C/H]$_{\rm gas}$\,=\,$x_{\rm C^+}$\,+\,$x_{\rm C^0}$\,+\,$x_{\rm CO}$\,+\,...\,, and its depletion factors  are essential parameters for understanding the gas and solid compositions that are ultimately incorporated into (exo)planets. The majority of protoplanetary disks are born in clusters  and, as a result, are  exposed to external far-ultraviolet (FUV)   radiation. These FUV photons potentially affect the disk's evolution, chemical composition,
and line excitation. We present the first detection of the  [\CI]\,609\,$\mu$m fine-structure (\mbox{$^3$P$_1$--$^3$P$_0$}) line of neutral carbon (C$^0$), achieved with ALMA, toward 
one of these disks, \mbox{d203-506}, in the Orion Nebula Cluster. We also report 
the detection of    [\CI] forbidden and \CI~permitted lines (from electronically excited energy levels up to $\sim$\,10~eV) observed with JWST in the near-infrared (NIR). These lines trace the irradiated outer disk and \mbox{photo-evaporative wind}. {Contrary to the common belief that these NIR lines are C$^+$ recombination lines, we find that they}  are {dominated}  by FUV-pumping of C$^0$ followed by fluorescence cascades. {They trace the transition from atomic to molecular gas,} and their intensities scale with $G_0$. 
The lack of outstanding   NIR \OI~fluorescent emission, however,  implies a sharper attenuation of external FUV radiation with \mbox{$E$\,$\gtrsim$\,12\,eV} ($\lambda$\,$\lesssim$\,Lyman-$\beta$). 
This is related to a lower {effective} FUV dust absorption cross section compared to that of interstellar grains, implying a more prominent role for FUV shielding by the C$^0$ photoionization continuum. 
The [\CI]\,609\,$\mu$m  line intensity is proportional to $N$(C$^0$) and can be used to infer  $x_{\rm C}$. We derive \mbox{$x_{\rm C}$\,$\simeq$\,1.4\,$\times$\,10$^{-4}$}. This  implies that there is no major depletion of volatile carbon compared to $x_{\rm C}$ measured in the natal  cloud, hinting at a young disk. We also show that external FUV radiation impacts the outer disk and wind by vertically shifting the water freeze-out depth, which likely results in less efficient grain growth and settling. This shift leads to nearly solar gas-phase C/O abundance ratios in these irradiated layers.}

\maketitle
%

\section{Introduction}\label{sec:introduction}

Tracing the carbon reservoir in protoplanetary disks is essential for understanding the variations in the gaseous elemental \mbox{carbon-to-oxygen} 
abundance\footnote{We denote the total ``solar'' abundance of element X   as $x_{\rm X_\odot}$, with $x_{\rm C_\odot}$\,=\,2.9\,$\times$\,10$^{-4}$ and $x_{\rm C_\odot}/x_{\rm O_\odot}$\,=\,0.54 \citep{Asplund09}.
We refer to the gas-phase  abundance of species Y  with respect to H nuclei as the density  ratio $x_{\rm Y}$\,=\,$n$(Y)/$n_{\rm H}$, where 
$n_{\rm H}$\,=\,$n$(H)\,+\,2$n$(H$_2$) in neutral gas.
}  ratio, which  connects the \mbox{atmospheric} composition of a planet with its formation site \citep[e.g.,][]{Oberg11,Madhusudhan12,Oberg23,Miotello19,Bosman21,Mah23}.
In addition, carbon plays a key role in the formation of the complex organics  that serve as the foundation for \mbox{prebiotic} chemistry and can be incorporated into planetary systems as they form.

\mbox{Observations} show that the gas-phase CO abundance in \mbox{1-10\,M\,yr} old disks can
be depleted by a factor of 10--100 compared to molecular clouds, even after accounting
for \mbox{freeze-out} and photodissociation. This suggests that something happens to CO by $\sim$1\,Myr \citep{Zhang20,Bergner20}. 
Grain growth, facilitated by the formation of water-rich ice mantles, followed by settling to the midplane and radial drift may explain these depletion factors
\citep[][]{Bergin16,Krijt20}. 
\begin{figure*}[h]
\centering   
\vspace{-0.45cm}
\includegraphics[scale=0.5, angle=0]{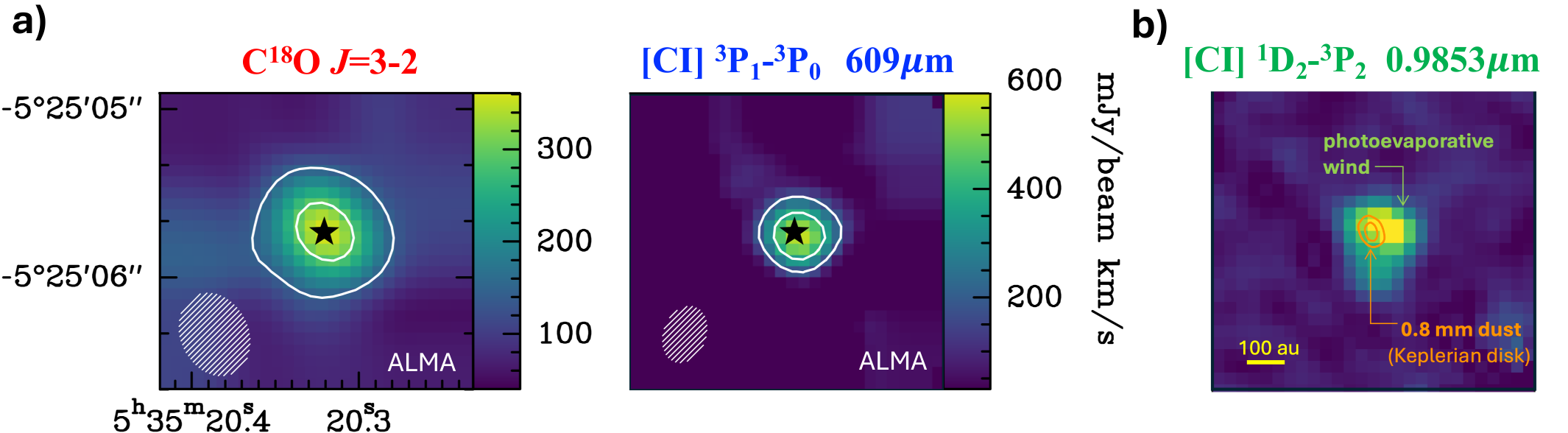}
\caption{Images of d203-506 (same field of view). 
\textit{a):} \mbox{C$^{18}$O\,$J$\,=\,3--2}
and [\CI]\,609\,$\mu$m integrated line emission (spatially unresolved). Contours show 
10$\sigma$ and 20$\sigma$ intensity levels for C$^{18}$O, and 5$\sigma$ and 10$\sigma$
for \mbox{[\CI]\,609\,$\mu$m}.
 Synthesized beam sizes are shown at the bottom corner.
 The apparent larger extent of the C$^{18}$O emission is due to the
 larger  beam and faint background emission.
 \textit{b):}  [\CI]\,0.9853\,$\mu$m emission from the photo-evaporative wind observed with NIRSpec 
 {(colored image).  Orange contours show
 the 0.8~mm dust continuum observed by ALMA  \citep[from][]{Berne24}.}}
\label{fig:d203-506_images}
\end{figure*}
However, the majority of observational studies target \mbox{``isolated''} disks in nearby \mbox{low-mass} star-forming regions such as 
Taurus, Lupus, Ophiuchus, and Chameleon \citep[at $d$\,$\sim$150\,pc; e.g.,][]{ALMA15,Oberg21}. 
Yet, most low-mass stars (and thus \mbox{planets}) are born in clusters 
 that contain one or more high-mass stars \mbox{\citep[e.g.,][]{Lada03}}. 
These \mbox{OB} stars emit intense ultraviolet (UV) radiation that illuminates the disk populations that gradually emerge from the natal  cloud \mbox{\citep[e.g.,][]{ODell93}}. \mbox{External} UV illumination leads to disk mass-loss driven by \mbox{photo-evaporation} 
\citep[][]{Johnstone98,Stoerzer99,Winter22} and potentially impacts the
 properties of planets forming within those disks \mbox{\citep{Winter22etal,Qiao23}}.
However, due to the greater distance to clusters, spectroscopic studies are scarcer {\citep[][]{Boyden20,Boyden23,Mauco23}}, and
it is still a matter of debate whether environmental UV radiation  modifies the chemistry of ``externally irradiated'' disks 
\citep[][]{Walsh13,Ramirez23,Javiera24}.

The \mbox{d203-506} system, in the Orion Nebula \mbox{Cluster} 
\citep[ONC; at $d$\,$\sim$\,400\,pc; see][]{Habart24},  is a remarkable
example of a disk
around a low-mass star that is externally irradiated  by far-ultraviolet (FUV) radiation
(\mbox{$\sim$6\,$<$\,$E$\,$<$\,13.6 eV}) with no evidence 
for the comet-shaped ionization front   seen in \mbox{proplyds}  \citep[e.g.,][]{ODell93,Ricci08}.
This implies a shielding from ionizing 
  extreme-ultraviolet (EUV) radiation (\mbox{$E$\,$>$\,13.6\,eV}).
The estimated FUV flux from the O-type stars \mbox{$\theta^1$\,Ori\,C} and \mbox{$\theta^2$\,Ori\,A}  is $\sim$\,10$^4$ times higher than  around the profusely studied isolated disks in Taurus-like clouds.  Thus, \mbox{d203-506}  may be more representative of the initial  conditions of the proto-Solar \mbox{System} disk \citep[][]{Fatuzzo08,Adams10,Bergin23}.
  This flared and nearly edge-on disk, located  toward the  
   Orion Bar \citep[][but see our \mbox{Appendix~\ref{app-location}}]{Champion17},   was first detected by the \textit{Hubble} Space Telescope in silhouette against the optical \mbox{nebular background
\citep{Bally00}}.  More recent
Keck, \textit{James Webb} Space Telescope (JWST),  and Very Large Telescope (VLT) Multi Unit Spectroscopic Explorer (MUSE) observations unveiled the extended nature of the 
 \mbox{FUV-pumped} \mbox{vibrationally} excited  H$_2$ \mbox{(hereafter H$_{2}^{*}$)} 
 and [\CI]\,0.8729\,$\mu$m emissions \citep{Habart23a,Haworth23,Berne24}.

The JWST PDRs4All team constrained  the mass of the host star
($\sim$0.3\,M$_{\odot}$), the disk's mass (\mbox{$\sim$\,10\,M$_{\rm Jup}$\,$\simeq$\,10$^{-2}$\,M$_{\odot}$}) and  radius \mbox{($\sim$100\,au\,$\simeq$\,0.25$''$)}, and a high  mass-loss rate
due to external \mbox{photo-evaporation} of \mbox{(0.1--4.6)\,$\times$\,10$^{-6}$\,M$_{\odot}$\,yr$^{-1}$} \citep{Berne24}.
The infrared (IR) spectrum shows the presence of highly excited rotational  lines of OH 
probing the  ongoing photodissociation of a hidden reservoir of water vapor \mbox{\citep{Zannese24}}. It also shows bright CH$^+$ and CH$_{3}^{+}$ rovibrational emission (which was first detected in space toward this disk by \citealt{Berne23} and \citealt{Changala23}). 
These ions form due to the high temperatures and the enhanced reactivity of H$_{2}^{*}$ with C$^+$ ions.
 Their abundances  scale with the FUV flux
\mbox{\citep[][]{Goico19}}.  These photochemical signatures differ from what is commonly observed in isolated disks 
\citep[e.g.,][]{vD23,Kamp23,Perotti23,Grant23}, suggesting that the outer disk gas reservoir
in \mbox{d203-506} is significantly reprocessed by
the external FUV field.

A powerful tracer of this  photo-processing  is the ground-state \mbox{$^3$P$_1$\,--\,$^3$P$_0$} fine-structure line  of neutral carbon  (C$^0$), the [\CI]\,609\,$\mu$m line, which can be observed by the Atacama Large Millimeter/submillimeter Array (ALMA) at \mbox{subarcsecond} and \mbox{sub-km\,s$^{-1}$} resolutions.
 However, low-angular-resolution searches toward  proplyds in the NGC\,1977 cluster 
 (\mbox{illuminated} by the B1V-type star 43 Ori, north of the ONC)  yielded 
non-detections, which were interpreted as being due to either very low-mass disks or a depletion of carbon in the outer disk \citep{Haworth22}. Here we present its first detection  toward an externally irradiated disk, complemented with 
the JWST detection of electronically excited C$^0$   lines in the near-infrared (NIR).

\section{ALMA and JWST observations of d203-506}\label{sec:observations}

We used ALMA to observe the protoplanetary disk \mbox{d203-506}, located at
\mbox{$\alpha$(2000)\,=\,5$^{\rm h}$35$^{\rm m}$20.32$^{\rm s}$},
 \mbox{$\delta$(2000)\,=\,$-$5$^{\rm \circ}$25$'$05.55$''$}.
These observations, a $\sim$40$''$$\times$40$''$ mosaic  using 47 ALMA \mbox{12 m} antennas, are part of an imaging program
of the Orion Bar (\mbox{2021.1.01369.S}, \mbox{P.I.: J. R. Goicoechea}). Here we present the detection of the [\CI]\,609\,$\mu$m
(492\,GHz, in band~8) and  \mbox{C$^{18}$O~$J$\,=\,3--2} (329\,GHz, in band~7) lines.  We used  correlators that provide $\sim$282\,kHz and $\sim$564\,kHz resolution, respectively. 
We binned all spectra to a common velocity resolution of 0.4\,km\,s$^{-1}$.
The total
observation times with the ALMA\,12\,m array 
were $\sim$4.6\,h (492\,GHz) and 2.7\,h (329\,GHz). 
The final synthesized beams are \mbox{0.52$''$\,$\times$\,0.38$''$ at
position angle PA\,=\,110$^{\circ}$} (492\,GHz)
and \mbox{0.77$''$\,$\times$\,0.60$''$ at PA\,=\,64$^{\circ}$} (329.3\,GHz). 
The  achieved  rms noises are 25\,mJy and 10\,mJy  per velocity channel, respectively.
 The complete data cubes and calibration strategy will be described
in an accompanying paper that focuses on the Bar. 

We also observed \mbox{d203-506}  as part of the  PDRs4All JWST Early Release Science program \citep[\mbox{ID \# 1288};][]{Berne22}.  
Our study focuses on NIRSpec's spectral cube from 0.97 to $\sim$5\,$\mu$m,  observed with grating dispersers at $R=\lambda/\Delta\lambda\sim2700$ resolution.
The angular resolution is $\sim$0.1$''$ ($\sim$40\,au at the distance to Orion).
\citet{Peeters24} \mbox{describe} the data reduction in detail.
 We extracted a NIR spectrum  in two apertures: one toward \mbox{d203-506} (ON measurement) and the other near the disk, 
providing the OFF reference.  The intrinsic spectrum of \mbox{d203-506}  is the {\mbox{ON\,--\,$f$$\cdot$\,OFF}} measurement (see \mbox{Appendices~\ref{app-distribution} and \ref{app-spectrum extraction}}).

\section{Results}\label{sec:results}

Figure~\ref{fig:d203-506_images} shows the detection of   
\mbox{[\CI]\,609\,$\mu$m} 
emission in \mbox{d203-506} as well as extended [\CI]\,0.9853\,$\mu$m forbidden line emission,
which connects the first electronic excited state $^1$D$_2$ with the ground state. 
To our knowledge, this is the first detection of these lines toward an externally photoevaporating disk\footnote{Nearby low-mass isolated disks such as DM Tau or TW Hya as well as Herbig Ae/Be disks show faint [\CI]\,609\,$\mu$m line
emission  \citep[][]{Tsukagoshi15,Kama16,Sturm22,Alarcon22,Pascucci23,Temmink23}.
 This  emission is driven  by internal FUV illumination
from the  central star. In addition, T\,Tauri stars show NIR \CI~permitted lines
(interpreted as \mbox{C$^+$\,+\,e$^-$} recombination lines) from the innermost $<$\,1\,au disk regions, interior to the dust sublimation radius
 \citep{McClure19,McClure20}. Interestingly, these inner regions  do not show
 lower-energy  [\CI] forbidden line emission at 0.9827 and 0.9853\,$\mu$m, which is bright in \mbox{d203-506} and in interstellar PDRs.}.
Only the NIRSpec [\CI]\,0.9853\,$\mu$m observation
 spatially resolves the outer disk and photo-evaporative wind.
 The  [\CI]\,0.9853\,$\mu$m emission 
 is similar to that of H$_{2}^{*}$ (see \mbox{Fig.~\ref{Fig:image_only_NIRSpec})}, 
 and significantly more extended than the 
  submillimeter dust continuum emission from the inner, dense Keplerian disk 
 (orange contours in \mbox{Fig.~\ref{fig:d203-506_images}b}).

\begin{figure}[t]
\centering    
 \includegraphics[scale=0.42, angle=0]{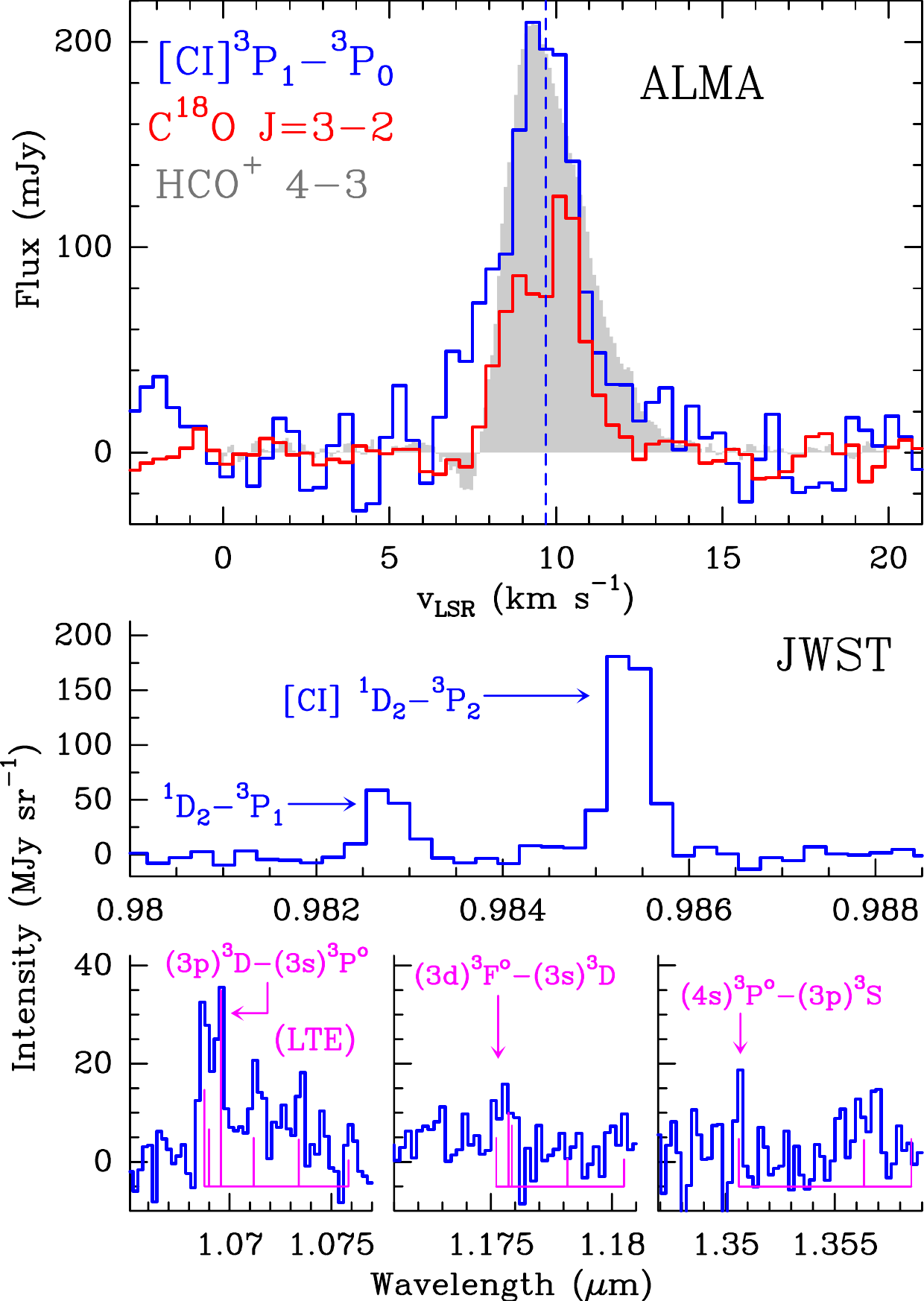}
\caption{Neutral carbon toward \mbox{d203-506}.
\textit{Upper:}  [\CI]\,609\,$\mu$m and  
\mbox{C$^{18}$O $J$\,=\,3--2} 
line profiles. The gray-shaded spectrum shows \mbox{HCO$^+$ $J$\,=\,4--3}
\citep{Berne24} scaled and convolved  to the angular resolution of the [\CI]\,609\,$\mu$m observations.
\textit{Lower:} NIR detection of [\CI] forbidden lines at 0.9827 and  0.9853\,$\mu$m (top) and \CI~permitted line
multiplets at $\sim$1.069, $\sim$1.176, and $\sim$1.355\,$\mu$m (bottom). Magenta lines
represent the position and  relative local thermodynamic equilibrium (LTE) intensity of each component.} 
\label{fig:d203-506_spectra}
\end{figure}

 \mbox{Figure~\ref{fig:d203-506_spectra}} shows the velocity-resolved \mbox{[\CI]\,609\,$\mu$m}
 and \mbox{C$^{18}$O 3--2}  line profiles observed by ALMA. 
The \mbox{[\CI]\,609\,$\mu$m} line does not show
the characteristic double-peak profile of an edge-on rotating  disk. Instead, it is
 similar to that of \mbox{HCO$^+$ 4--3} (convolved to the same angular resolution), which is
 dominated by emission from the outer disk  and the quasi-spherical photo-evaporative wind
\citep{Berne24}. Since our observation does not spatially resolve \mbox{d203-506}, the resulting line profile is more or less Gaussian. The remaining  blueshifted \mbox{[\CI]\,609\,$\mu$m} emission
may originate from  {a slow} wind not
traced by HCO$^+$. {A two-Gaussian fit to the two  \mbox{[\CI]\,609\,$\mu$m} components shows
a velocity centroid difference of only 2.3\,$\pm$\,0.4\,km\,s$^{-1}$, which is inconsistent with emission from the fast jet that emerges from the object} \citep[e.g.,][]{Haworth23}.
In contrast, \mbox{C$^{18}$O 3--2} shows a {clear} double-peak profile, which traces
deeper layers (we infer \mbox{$N_{\rm H}$\,$\gtrsim$\,10$^{23}$\,cm$^{-2}$}) of the Keplerian disk \citep[][]{Miotello16}.

 Figure~\ref{fig:d203-506_spectra} also shows NIR excited lines from neutral carbon:
 [\CI]  \mbox{(2$p^2$)\,$^1$D$_2$\,--\,(2$p^2$)\,$^3$P}  \mbox{forbidden}  lines at  0.9827 and  0.9853\,$\mu$m (\mbox{$E_{\rm u}$\,$\simeq$\,1.3\,eV}), and the highly excited 
\mbox{(3$p$)\,$^3$D\,--\,(3$s$)\,$^3$P$^{\rm o}$}, \mbox{(3$d$)\,$^3$F$^{\rm o}$\,--\,(3$p$)\,$^3$D},
and \mbox{(4$s$)\,$^3$P$^{\rm o}$\,--\,(3$p$)\,$^3$S}
\mbox{permitted} line multiplets at $\sim$1.069, $\sim$1.176, and $\sim$1.355\,$\mu$m, respectively (with \mbox{$E_{\rm u}$\,$\simeq$\,10\,eV}). 
In addition, \cite{Haworth23} imaged the [\CI] \mbox{(2$p$)\,$^1S_0$\,--\,(2$p$)\,$^1D_2$} 
forbidden line emission
at 0.8729\,$\mu$m (\mbox{$E_{\rm u}$\,$\simeq$\,2.7\,eV}) with VLT/MUSE.
Table~\ref{Table_intensities} summarizes the observed line intensities.
As we show here, all these lines arise from  neutral  photon-dominated region (PDR) gas (not from H$^+$ ionized gas), which is consistent with the low ionization potential (IP) of C$^0$ (\mbox{11.26\,eV}) and
with the lack of cometary ionization fronts in \mbox{d203-506}. 

\section{Analysis: Thermochemistry and line excitation}
\label{sec:analysis}

Neutral carbon lines are expected to trace warm molecular gas at the \mbox{C$^+$/C$^0$/CO} transition zone of the outer disk and inner photo-evaporative  wind \mbox{\cite[e.g.,][]{Haworth20}}. At the high gas densities relative to the FUV
flux in a disk (\mbox{$n_{\rm H}/G_0$\,$>$\,10$^2$\,cm$^{-3}$}), H$_2$ and CO line \mbox{self-shielding} shifts  the atomic to molecular gas transitions close to the irradiated surface, where the FUV flux is strong and the \mbox{H$_2$-emitting} gas is heated
to $T_{\rm k}$\,$\gtrsim$\,1000\,K \citep[][]{Johnstone98,Stoerzer99,Champion17}. To derive the beam-averaged column density of C$^0$  in this  zone, $N$(C$^0$),
and the (total) gas-phase  abundance of carbon,
\mbox{$x_{\rm C}$\,=\,$x_{\rm C^+}$\,+\,$x_{\rm C^0}$\,+\,$x_{\rm CO}$\,+\,...\,},
 we used the Meudon code \mbox{\citep{LePetit06}},  a {fully benchmarked   PDR model \citep[see][]{Rollig07}}.
 We simulated the [\CI]\,609\,$\mu$m- and NIR-emitting zones as a 1D stationary slab of constant-density gas. This model involves {solving the $\lambda$-dependent} attenuation of external FUV photons \citep[considering dust extinction and line self-shielding;][]{Goico07}, C$^0$ photoionization, C$^+$ recombination, and  CO photodissociation.
{Our detailed treatment of the penetration of FUV radiation, H$_2$ excitation,  thermal balance, and chemistry 
allows a precise determination of the H/H$_2$ and C$^+$/C/CO transition layers in the slab,
but neglects the wind \mbox{dynamics} \citep[e.g.,][]{Haworth20}.} 
We implicitly assumed that the contribution to the [\CI]\,609\,$\mu$m intensity from the inner disk close to the central star \citep[][]{Gressel20} is negligible and beam-diluted.
We treated the $\lambda$-dependent  absorption and \mbox{scattering} of  
FUV photons  by dust, providing an {\mbox{effective}} absorption cross section of \mbox{7$\times$10$^{-22}$\,cm$^{2}$\,H$^{-1}$}  at 1000\,\AA. These 
grains  {are bigger} than interstellar medium (ISM) grains but
consistent with the {modest grain growth}  expected in the upper layers of irradiated disks \mbox{\citep{Stoerzer99,Birnstiel18}}.

\begin{figure}[t]
\centering   
\includegraphics[scale=0.42, angle=0]{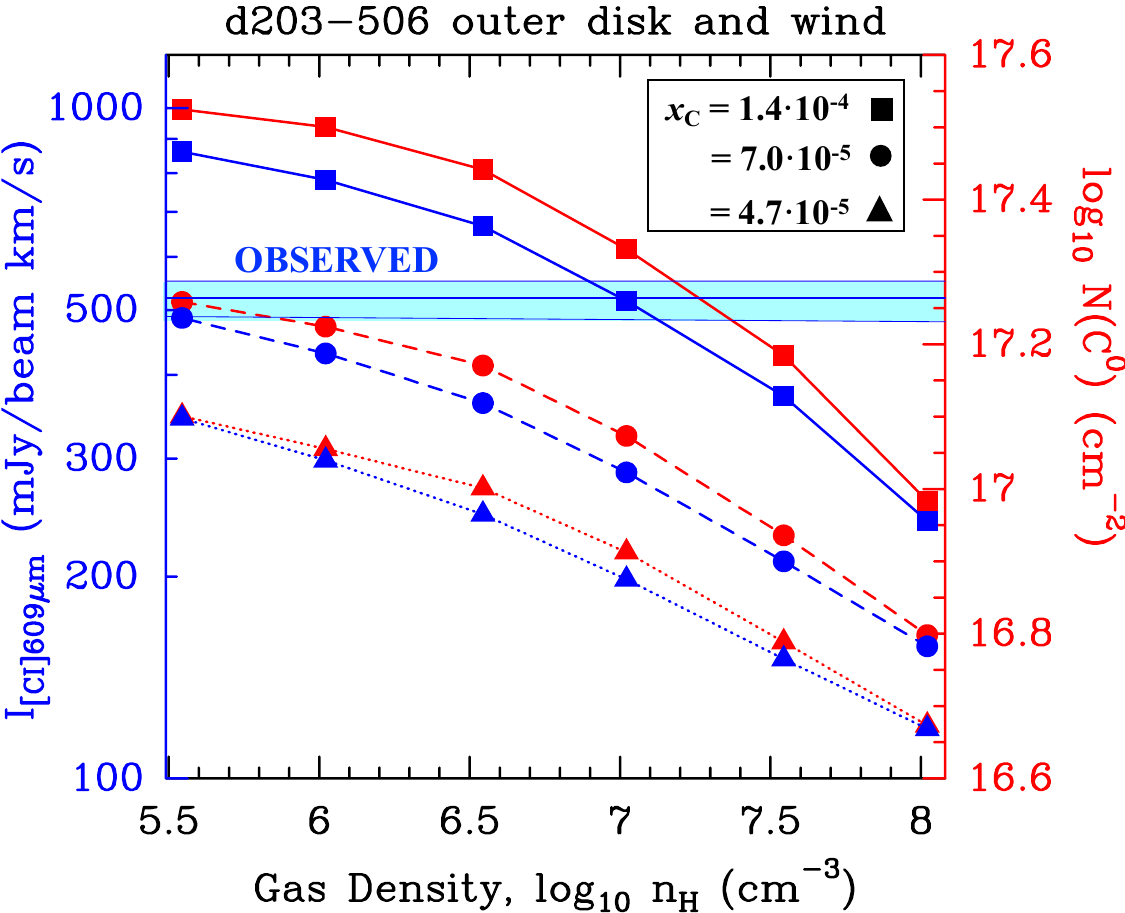}
\caption{Models with \mbox{$G_0$\,$=$2$\times$10$^4$} and different values
of $n_{\rm H}$ and $x_{\rm C}$ . Red and blue markers  show the predicted C$^0$ column density
and  [\CI]\,609\,$\mu$m intensity, respectively, from
$A_V$\,=\,0 to 10 mag into the wind and disk system.
The gas density  derived from H$_2$ observations is
$\approx$\,10$^7$\,cm$^{-3}$ \citep{Berne24}. The horizontal line marks the observed intensity ($\pm$\,1$\sigma$). Circles and triangles represent models with small depletion factors of two and three, respectively.}
\label{fig:PDR_d203-506-grid}
\end{figure}

The [\CI]\,609 and 370\,$\mu$m  lines involve forbidden transitions 
with very low Einstein coefficients for spontaneous emission. 
In dense and warm gas, and because of their low excitation requirements --- low \mbox{critical density} 
 (\mbox{$n_{\rm cr}$\,$\lesssim$\,10$^3$\,cm$^{-3}$})
and  low level energy separation compared to $T_{\rm k}$  
\mbox{($\Delta E$/$k_{\rm B}$\,=\,23.6\,K\,$\ll$\,$T_{\rm k}$) ---} the 
[\CI]\,609\,$\mu$m  emission is optically 
thin\footnote{At high temperatures, $T_{\rm k}$\,$>$\,500--1000\,K, the [\CI]\,609\,$\mu$m line becomes
optically thick only if $N$(C$^0$) is greater than several times 10$^{19}$\,cm$^{-2}$.}, collisionally excited, and 
thermalized ($T_{\rm ex}$\,=\,$T_{\rm k}$).
In this regime,  
\mbox{$I$([\CI]\,609\,$\mu$m)} is proportional
to $N$(C$^0$)  irrespective of the physical conditions. In addition, 
 $N$(C$^0$) is not very sensitive to 
variations in $G_0$ (\mbox{Appendix~\ref{app-atomic_carbon}}).
\mbox{Figure~\ref{fig:PDR_d203-506-grid}} shows the $N$(C$^0$) and \mbox{$I$([\CI]\,609\,$\mu$m)}, integrating up to an equivalent extinction depth of $A_V$\,=\,10\,mag,  predicted by  models of varying gas densities, with  \mbox{$n_{\rm H}$\,=\,$n$(H)\,+\,2$n$(H$_2$)},
and  \mbox{$G_0$\,=\,2\,$\times$\,10$^4$}. 
Our models encompass the [\CI]\,609\,$\mu$m-emitting layers and thus can be used, for a given $n_{\rm H}$, to constrain $N$(C$^0$) and 
\mbox{$x_{\rm C}$}.
 \mbox{Figure~\ref{fig:PDR_d203-506-grid}} shows that \mbox{$I$([\CI]\,609\,$\mu$m)},
a surface tracer, scales with $N$(C$^0$) and
$x_{\rm C}$. 
Models marked with squares have \mbox{$x_{\rm C}$\,=\,1.4\,$\times$\,10$^{-4}$}, which is the gas-phase abundance of interstellar
carbon measured\footnote{From UV absorption-line measurements toward the $\theta^1$ Ori B star in the Trapezium. The gaseous  
$x_{\rm C}/x_{\rm O}$
ratio in this interstellar sight line is \mbox{0.52\,$\pm$\,0.18} \citep{Cartledge01}, 
very close to the solar ratio.
These values imply moderate depletions, about \mbox{$-$0.3~dex}  with respect to their solar abundances, consistent with their incorporation into 
refractory dust  \citep[e.g.,][]{Savage96,Jenkins09}.
Other studies suggest \mbox{$x_{\rm C}$\,$\simeq$\,2\,$\times$\,10$^{-4}$} in Orion's nebular gas
\mbox{\citep{Simon-Diaz11}} and B-type stars \citep{Nieva12}, with \mbox{$x_{\rm C}$\,/\,$x_{\rm O}$\,$\simeq$\,0.4--0.5}.} 
in the line of sight toward the Trapezium stars \citep{Sofia04}.
Our main result  is that the observed 
\mbox{$I$([\CI]\,609\,$\mu$m)} implies high $x_{\rm C}$ abundances, 
roughly the same as those in the natal cloud. 
The lack of significant volatile carbon depletion,
which would evolve with time \citep[e.g.,][]{Kama16,Krijt20}, suggests that this protoplanetary disk is young.
  
\begin{figure}[t]
\centering   
\includegraphics[scale=0.46, angle=0]{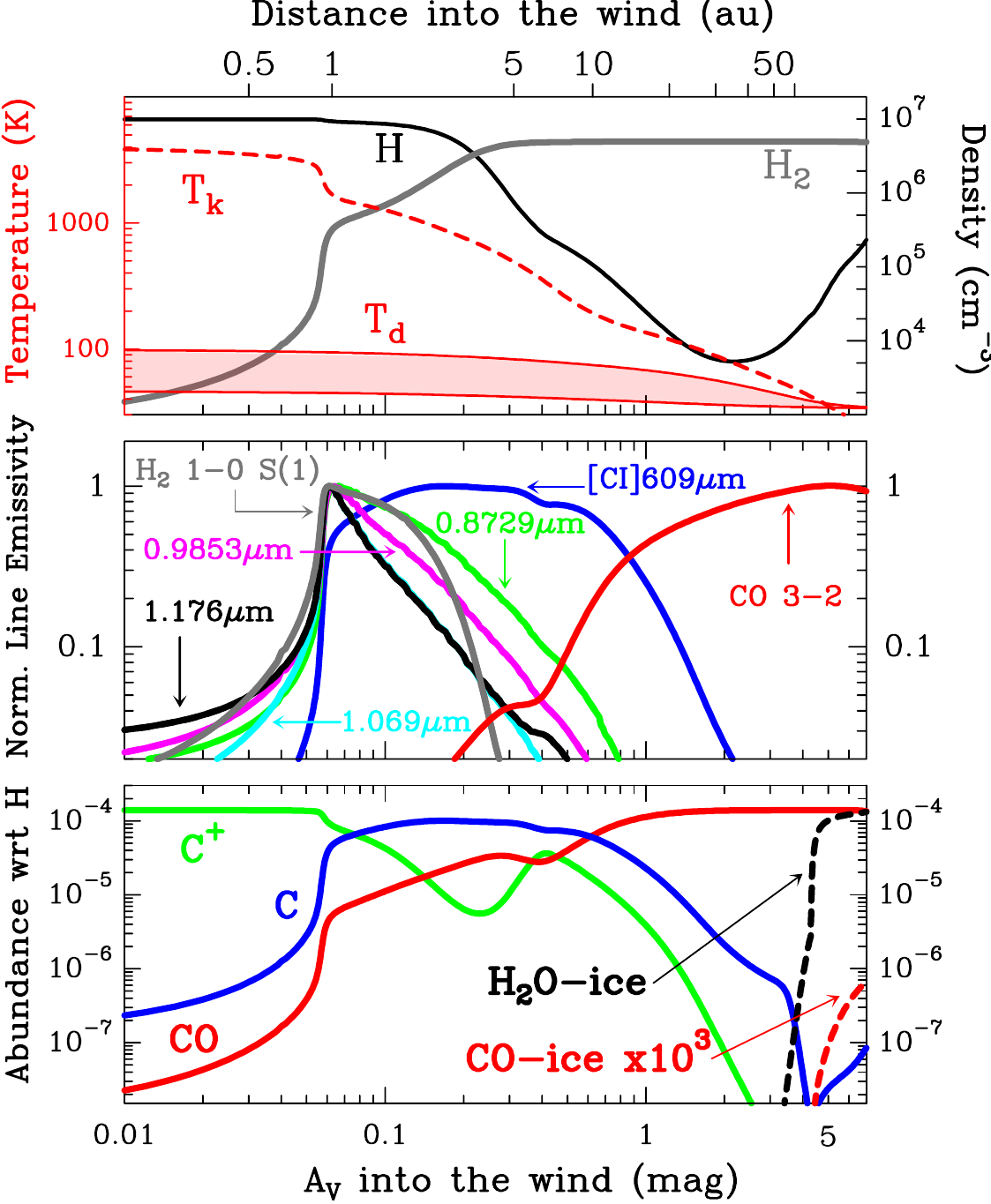}
\caption{Model of the outer disk and {inner}  wind  
(\mbox{$n_{\rm H}$\,=\,10$^7$\,cm$^{-3}$} and \mbox{$G_0$\,=\,2$\times$10$^4$}). \textit{Upper panel}: H$_2$ and H densities and gas and dust temperatures  as a function of depth into the disk. 
\textit{Middle}: Normalized line emissivity profiles, including NIR C$^0$ lines.
\textit{Lower}: Abundance profiles.} 
\label{fig:PDR_structure_203-506}
\end{figure}

We adopted \mbox{$n_{\rm H}$\,$\approx$\,10$^7$\,cm$^{-3}$}
as the gas  density in the \mbox{[\CI]\,609\,$\mu$m-emitting} zone,
as determined from H$_2$ observations with JWST \citep{Berne24}.  
At these densities, the spatial scale of the [\CI]\,609\,$\mu$m emission is a few tens of au,  a fraction of an arcsecond, in agreement  with the observed size of \mbox{d203-506}.
 Models  with  \mbox{$n_{\rm H}$\,$\simeq$\,10$^7$\,cm$^{-3}$} fit the observed
$I$([\CI]\,609\,$\mu$m)  with  
$N$(C$^0$)\,$\simeq$\,2$\times$10$^{17}$\,cm$^{-2}$
and  \mbox{$x_{\rm C}$\,$\simeq$\,1.4\,$\times$\,10$^{-4}$}, with an uncertainty of less
than a factor of 2. 
Figure~\ref{fig:PDR_structure_203-506} (upper panel) shows the predicted structure of
the [\CI]\,609\,$\mu$m-emitting layer, roughly representing the vertical structure of the outer disk
and inner wind. 
The \mbox{middle} panel shows the normalized 
\mbox{[\CI]\,609\,$\mu$m}, \mbox{H$_2$ 1--0 $S$(1)}, and
\mbox{CO 3--2} line emissivities. 
 \mbox{[\CI]\,609\,$\mu$m} peaks  just beyond the H$_2$ dissociation front.
{The lower panel  shows the C$^+$, C$^0$, CO, water ice,  and CO ice abundance profiles}.
The predicted gas temperature in the H$_{2}$-emitting zone is \mbox{$T_{\rm k}$\,$\gtrsim$\,1000--1200\,K}, 
consistent with the observed H$_2$ rotational temperatures 
\citep{Berne24}, and drops to \mbox{$T_{\rm k}$\,$\simeq$\,700\,K} at the \mbox{[\CI]\,609\,$\mu$m} intensity  peak, where C$^0$ becomes the most abundant carbon species.  

Turning back to the NIR carbon lines, we note that the \mbox{(3$p$)\,$^3$D\,--\,(3$s$)\,$^3$P$^{\rm o}$} (1.069\,$\mu$m) 
and \mbox{(3$d$)\,$^3$F\,--\,(3$p$)\,$^3$D} (1.176\,$\mu$m)  multiplets are the brightest
permitted lines predicted by  recombination theory 
\citep[\mbox{C$^+$\,$+$\,e$^-$\,$\rightarrow$\,C$^*$\,+\,line cascade};][]{Escalante90}.
However,  the observed  \mbox{$I$(1.069\,$\mu$m)/$I$(1.176\,$\mu$m)}
and  \mbox{$I$(0.984\,$\mu$m)/$I$(1.069\,$\mu$m)} intensity ratios do not match the ratios predicted by this theory
\mbox{(see Appendix~\ref{app-recomb-theory})}. In analogy with the NIR  \OI~fluorescent lines detected very close
behind the ionization fronts of 
\mbox{interstellar} PDRs \citep[][]{Walmsley00,Lucy02,Peeters24}, the NIR C$^0$~lines could form via de-excitation cascades  following \mbox{FUV-pumping} from the ground-state $^3$P to the high-energy triplets 
(3$s$)\,$^3$P$^{\rm o}$, (3$d$)\,$^3$D$^{\rm o}$, (3$d$)\,$^3$F$^{\rm o}$, (4$s$)\,$^3$P$^{\rm o}$, and so on
 \mbox{(i.e., C\,$+$\,$h\nu_0$\,$\rightarrow$\,C$^*$\,+\,line cascade)}. The  \mbox{FUV-pumping}
lines of these transitions lie at 1656\,$\AA$, 1277\,$\AA$, 1279\,$\AA$, and 1280\,$\AA$, respectively,
that is, FUV photons with energies \mbox{$h\nu_0$\,=\,7.5--9.6\,eV}   
 \mbox{(see Fig.~\ref{Fig:energy_levels_all}} for a Gotrian diagram). 
Here we  included FUV-pumping followed by radiative cascades in the Meudon PDR code (Appendix~\ref{app-new-FUV-pumping-model}). This  mechanism  greatly increases the intensities
of all the observed NIR C$^0$ lines (both forbidden and permitted) compared to models that  consider collisional excitation and  C$^+$ recombination alone, thereby fitting the observations (\mbox{Table~\ref{Table_intensities}}).
 NIR C$^0$ line intensities scale with $G_0$ \mbox{(see Appendix~\ref{app-atomic_carbon})} and are a powerful diagnostic of {external}  FUV irradiation (we derive {\mbox{$G_0$\,$\simeq$\,(1--2)$\times$10$^4$}} in \mbox{d203-506}).
 \mbox{Figure~\ref{fig:PDR_structure_203-506}}  shows their   emissivity profiles, which peak very close to  the 
H$_2$ 1--0 $S(1)$ emission, as observed.
{Thus, NIR C$^0$ lines trace the transition from neutral atomic gas to molecular gas}. 
The lack of bright NIR \OI~fluorescent lines 
{(other than a  faint \OI~\,1.317\,$\mu$m emission;} Appendix~\ref{app-spectrum extraction})  
implies that \mbox{d203-506}  is pervaded by a reduced flux of FUV photons
at the pumping transitions of these lines:  photons
with  \mbox{$E$\,$\gtrsim$\,12\,eV} 
\citep[\mbox{$\lambda$\,$\lesssim$\,Lyman--$\beta$},][]{Walmsley00}.
Furthermore, grain sizes and gas-to-dust mass ratios in the outer disk and wind  are {likely} larger than those in the ISM \mbox{\citep[e.g.,][]{Winter22}}. This results in lower \mbox{{effective}} FUV dust absorption cross sections, and implies that the  attenuation of these FUV photons is dominated by  
{\HI~and H$_2$} absorption lines, and by the C$^0$ photoionization {\mbox{continuum}} at \mbox{$\lambda$\,$<$\,1101\,\AA}
(\mbox{Appendix~\ref{app-gas_absorption}}). In contrast, the attenuation of FUV photons at the pumping transitions of C$^0$ (\mbox{$\lambda$\,$\simeq$\,1300\,\AA}) is less pronounced and is dominated by dust grains.
The lack of outstanding oxygen fluorescent  lines but detectable 
carbon lines in \mbox{d203-506} agrees with this scenario.

\section{Discussion: Photo-processing in d203-506}
\label{sec:discussion}

The  \mbox{[\CI]\,609\,$\mu$m} line fluxes observed in nearby isolated disks (i.e., disks that are not externally irradiated) such as \mbox{TW Hya}   suggest carbon depletion by 1--2 orders of magnitude in these disks \mbox{\citep[e.g.,][]{Kama16}}. 
 In \mbox{d203-506}, however, we determine a
 high gas-phase $x_{\rm C}$ abundance. 
 Furthermore, the outer  layers of externally FUV-irradiated disks are heated to high gas and dust temperatures, $T_{\rm k}$ up to 1000 K and $T_{\rm d}$ up to $\simeq$\,100\,K
(red curves in the upper panel of Fig.~\ref{fig:PDR_structure_203-506}). These warm dust
temperatures are on the order of the freezing temperature of H$_2$O  
and  significantly higher than that of CO \citep[$\sim$\,25\,K; e.g.,][]{Oberg11},
thus preventing their freeze-out {in the outer disk and wind}. 

 A strong external FUV flux leads to high ice photo-desorption rates and high $T_{\rm d}$,
and shifts the freeze-out depth (e.g., in disk height) where most oxygen becomes trapped in  ice  mantles  (see \mbox{Fig.~\ref{fig:PDR_structure_203-506}} and \mbox{Appendix~\ref{app-snowline}}).
Thus, we expect the vertical distribution of the $x_{\rm C}$\,/$x_{\rm O}$ abundance ratio in irradiated disks to differ from that of isolated disks.
 In particular, {grain} growth, settling,
and subsequent radial drift \citep{Krijt20,Zhang20} will be less efficient due to the lack of thick ice mantles in the strongly irradiated outer disk. 
This will  result in  lower depletions of gaseous CO  compared to more
 evolved, $\gtrsim$\,1\,Myr, disks \mbox{\citep[e.g.,][]{Bergner20}}.
 {Interestingly, the NIRSpec spectrum of d203-506 shows bright CO ro-vibrational emission 
 at $\sim$4.7\,$\mu$m \citep[see][]{Berne24}, which we attribute to the externally
 irradiated layers.}
All in all, the outer gas layers of \mbox{d203-506}, and presumably of other  externally irradiated disks \citep[e.g.,][]{Boyden23}, must be \mbox{``oxygen rich,''}
with \mbox{$x_{\rm C}/x_{\rm O}$} very close to the solar ratio
(assuming gaseous oxygen abundances as in the Orion cloud$^4$).
In \mbox{d203-506}, this is {further} demonstrated by the presence of abundant  gaseous OH, with \mbox{$x_{\rm OH}+x_{\rm H_2O}$\,$\gtrsim$\,10$^{-5}$} \mbox{\citep{Zannese24}}, and
by the detection of bright and extended [\OI]\,6302\,\AA~emission  \citep[][]{Haworth23}, where the  \mbox{excited}
O($^1$D) atoms {in PDR  gas} (close to the H/H$_2$ dissociation front) are  produced by OH photodissociation \citep{EvD83,Storzer98b}.
In these irradiated layers, water undergoes photo-processing and cycles between gas-phase formation and photodissociation.  
Other \mbox{inherited} ice mantles will undergo \mbox{photo-desorption} and photo-processing as well. 
Does this mean that strongly irradiated disks develop a less rich carbon chemistry than isolated disks? On the contrary, we expect that hydrocarbons will be abundant in these   O-rich 
 disk layers, formed by gas-phase reactions triggered by the external FUV radiation, high temperatures, and enhanced abundances of C$^+$, C$^*$, C$^0$, and H$_{2}^{*}$ \citep[see also][]{Agundez08,Berne23}.
These FUV-induced processes, facilitated by vertical transport, likely lead to different chemical compositions --- similar to interstellar PDRs \citep[e.g.,~rich in hydrocarbons,][]{Cuadrado15} --- in the outer disk regions 
where {gas-giant planets  form}.
\mbox{Additionally}, the absence of carbon depletion in other proplyds would imply low disk masses for the non-detections in NGC~1977 by \cite{Haworth22}.
 Still, even higher angular resolution  will be needed to  map  radial and vertical
 chemical abundances in Orion's  disks.

\begin{acknowledgements}
We made used of ADS/JAO.ALMA\#2021.1.01369.S data. ALMA is a partnership of ESO (representing its member states), NSF (USA) and NINS (Japan), together with NRC (Canada), NSTC and ASIAA (Taiwan), and KASI (Republic of Korea), in cooperation with the Republic of Chile. The Joint ALMA Observatory is operated by ESO, AUI/NRAO and NAOJ.
The JWST data were obtained from the Mikulski Archive for Space Telescopes at the Space Telescope Science Institute, which is operated by the Association of Universities for Research in Astronomy, Inc., under NASA contract NAS 5-03127 for JWST. These observations are associated with program \#1288. Support for program \#1288 was provided by NASA through a grant from the Space Telescope Science Institute, which is operated by the Association of Universities for Research in Astronomy, Inc., under NASA contract NAS 5-03127.  We thank our referee for a constructive report.
JRG thanks the Spanish MCINN for funding support under grants
\mbox{PID2019-106110GB-I00} and \mbox{PID2023-146667NB-I00}.
TJH acknowledges funding from a Royal Society Dorothy Hodgkin Fellowship and UKRI guaranteed funding for a Horizon Europe ERC consolidator grant (EP/Y024710/1).
We thank 
the Programme National ``Physique et Chimie du Milieu Interstellaire'' (PCMI) of CNRS/INSU with INC/INP, co-funded by CEA and CNES.
EP acknowledges support from the University of Western Ontario, the Institute for Earth and Space Exploration, the Canadian Space Agency (CSA, 22JWGO1-16), and the Natural Sciences and Engineering Research Council of Canada. TO is supported by JSPS Bilateral Program, Grant Number 120219939.

\end{acknowledgements}

\bibliographystyle{aa}
\bibliography{references}

\begin{appendix}\label{Sect:Appendix}

\section{On the location of \mbox{d203-506}}
\label{app-location}

\begin{figure}[h]
\centering   
\includegraphics[scale=0.35, angle=0]{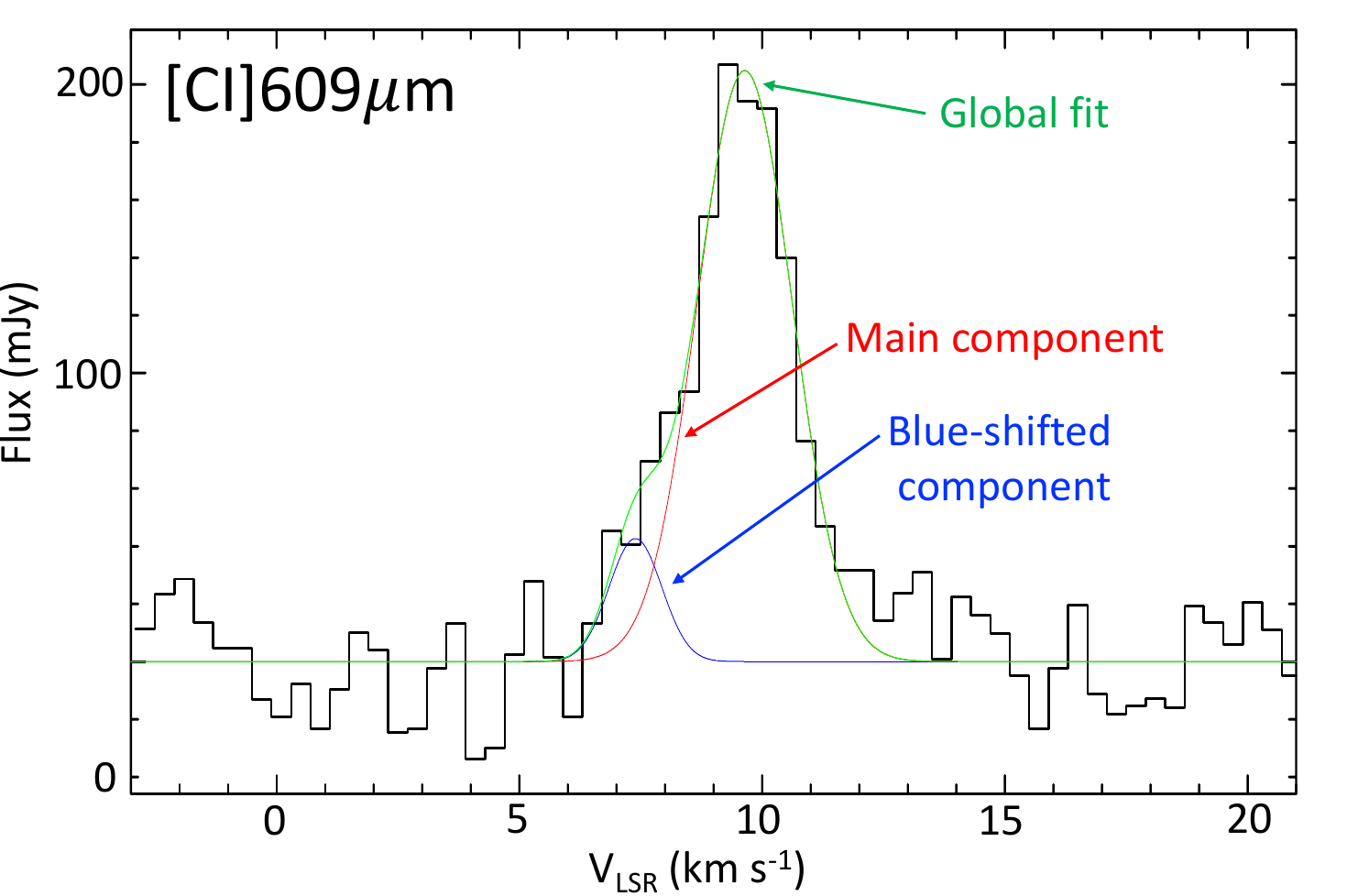}
\caption{{[\CI]\,609\,$\mu$m line profile 
toward \mbox{d203-506} and the two-component Gaussian fit. The velocity centroid difference
of the two components is \mbox{2.3\,$\pm$\,0.4\,km\,s$^{-1}$} and suggests
that the blueshifted emission arises from a slow wind not seen in HCO$^+$ 4--3
emission (see Fig.~\ref{fig:d203-506_spectra})}.}
\label{Fig:Gauss_fit}
\end{figure}

The NIR [\CI] lines observed by JWST in \mbox{d203-506} are \mbox{3--5} times brighter  than toward the
Orion Bar PDR \mbox{\citep{Peeters24}}.
The submm \mbox{[\CI]\,609\,$\mu$m (\mbox{$^3$P$_1$--$^3$P$_0$})}  line integrated flux  is 605\,mJy\,km\,s$^{-1}$, slightly above the CO~3-2 line fluxes
measured by \cite{Boyden20,Boyden23} in other disks of the ONC.
The velocity centroid of the main \mbox{[\CI]\,609\,$\mu$m} line emission,
\mbox{$v_{\rm LSR}$\,(\CI)\,=\,9.5\,$\pm$\,0.1\,km\,s$^{-1}$}
{(see Fig.~\ref{Fig:Gauss_fit})},
is in the range of LSR velocities of these disks. 
This velocity would be consistent with 
\mbox{d203-506} being embedded at the rim of the Orion Bar where
only FUV photons with $E<13.6$\,eV  penetrate inside.
The amount of extinction  internal to the Bar PDR can be estimated using the observed line intensity ratio of H$_2$ lines  that arise from the same upper $v$ and $J$ state. 
The inferred extinction 
toward  \mbox{d203-506}  is nearly zero \citep[Fig.~11 of][]{Peeters24}, suggesting that 
the disk is actually not embedded in the Bar.
\mbox{Indeed}, \mbox{d203-506} shows a silhouette appearance {at the wavelengths
of several visible and NIR  lines} (H$\alpha$, [\SII]\,6732\,\AA, {\mbox{Paschen-$\alpha$}}, etc.)  arising from ionized gas in the background \HII~region. This lead \cite{Haworth23} to conclude  that the disk lies  in front  of  a structure of ionized gas known as the near ionization layer \citep[NIL,][]{Abel19,ODell20}. Thus, located closer to the observer than
the Bar, the \mbox{Huygens} \HII~region, and the Trapezium stars.
In this picture, \mbox{d203-506} is in the foreground,  well separated from the  Bar, at least by $\sim$0.4\,pc.
The NIL would absorb any EUV radiation  emitted by the massive stars \mbox{$\theta^1$ Ori C} (in the \mbox{Trapezium}) and \mbox{$\theta^2$ Ori A}, thus leaving the disk illuminated only by FUV radiation. 
 Still, the  LSR velocity
of the ionized gas emission in the NIL, $-12$\,$\pm$\,2\,km\,s$^{-1}$ \citep{ODell20} is very different to that of \mbox{d203-506}, so this layer is likely not related with the birth place of the disk. 
Yet, the strong external FUV field and high gas-phase carbon abundance derived in \mbox{d203-506} is very similar to that in the Orion cloud, suggesting a young disk, but  separated from the main star-forming molecular \mbox{core (OMC-1)}.
Plausible scenarios are: i) the disk has recently emerged from a  molecular globule, such as those detected by \cite{Goico20} in the Veil bubble of Orion,
or, ii) \mbox{d203-506} belongs to a recent star formation episode followed by radial
migration  \mbox{\citep[e.g.,][]{Stoerzer99,Winter19}}.
Irrespective of its exact location, the important point is that \mbox{d203-506} is externally illuminated by  FUV radiation, but no EUV.

\section{Spatial distribution of the  \mbox{[\CI]\,0.9853\,$\mu$m} emission
in \mbox{d203-506}}
\label{app-distribution}

\begin{figure}[t]
\centering   
\includegraphics[scale=0.95, angle=0]{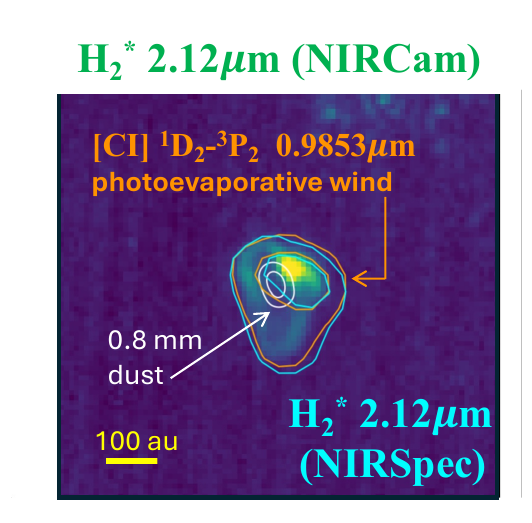}
\caption{[\CI]\,0.9853\,$\mu$m observed with NIRSpec (orange contours
 at \mbox{3$\times$10$^{-4}$} and       \mbox{6$\times$10$^{-4}$~erg\,s$^{-1}$\,cm$^{-2}$\,sr$^{-1}$)}
 over a NIRCam image of the H$_{2}^{*}$ 1--0 $S(1)$ emission.
 Cyan contours represent the \mbox{H$_{2}^{*}$ 1--0 $S(1)$} emission observed by
 NIRSpec  (7$\times$10$^{-4}$ and 2$\times$10$^{-3}$~erg\,s$^{-1}$\,cm$^{-2}$\,sr$^{-1}$)
 at a similar angular resolution and spatial sampling as the [\CI]\,0.9853\,$\mu$m emission.}
\label{Fig:image_only_NIRSpec}
\end{figure}

Figure~\ref{Fig:image_only_NIRSpec} compares the \mbox{[\CI]\,0.9853\,$\mu$m} (orange contours) and \mbox{H$_{2}^{*}$ 1--0 $S(1)$} (cyan contours) emission observed with NIRSpec at similar angular resolution and spatial sampling. These observations have worse resolution and sampling than the \mbox{H$_{2}^{*}$ 1--0 $S(1)$} image obtained by NIRCam (greenish colors). They show that both line emissions have very similar spatial distributions {(at the \mbox{$\sim$0.1$''$\,$\simeq$\,40\,au} resolution of these images)}. {Since in a PDR the intensity of FUV-pumped H$_{2}^{*}$ lines peaks slightly ahead of the H/H$_2$ transition, the observed NIR C$^0$ lines also trace this 
transition from neutral atomic to molecular gas (as confirmed by our model predictions in Fig.~\ref{fig:PDR_structure_203-506}), that is, the FUV-irradiated gas layers where the H$_2$ abundance sharply increases. We note that 
spatially resolving the H$_{2}^{*}$ and NIR [\CI] line emissivity gradients will require an order of magnitude higher angular resolution.}

\begin{figure*}[th]
\centering   
\includegraphics[scale=0.391, angle=0]{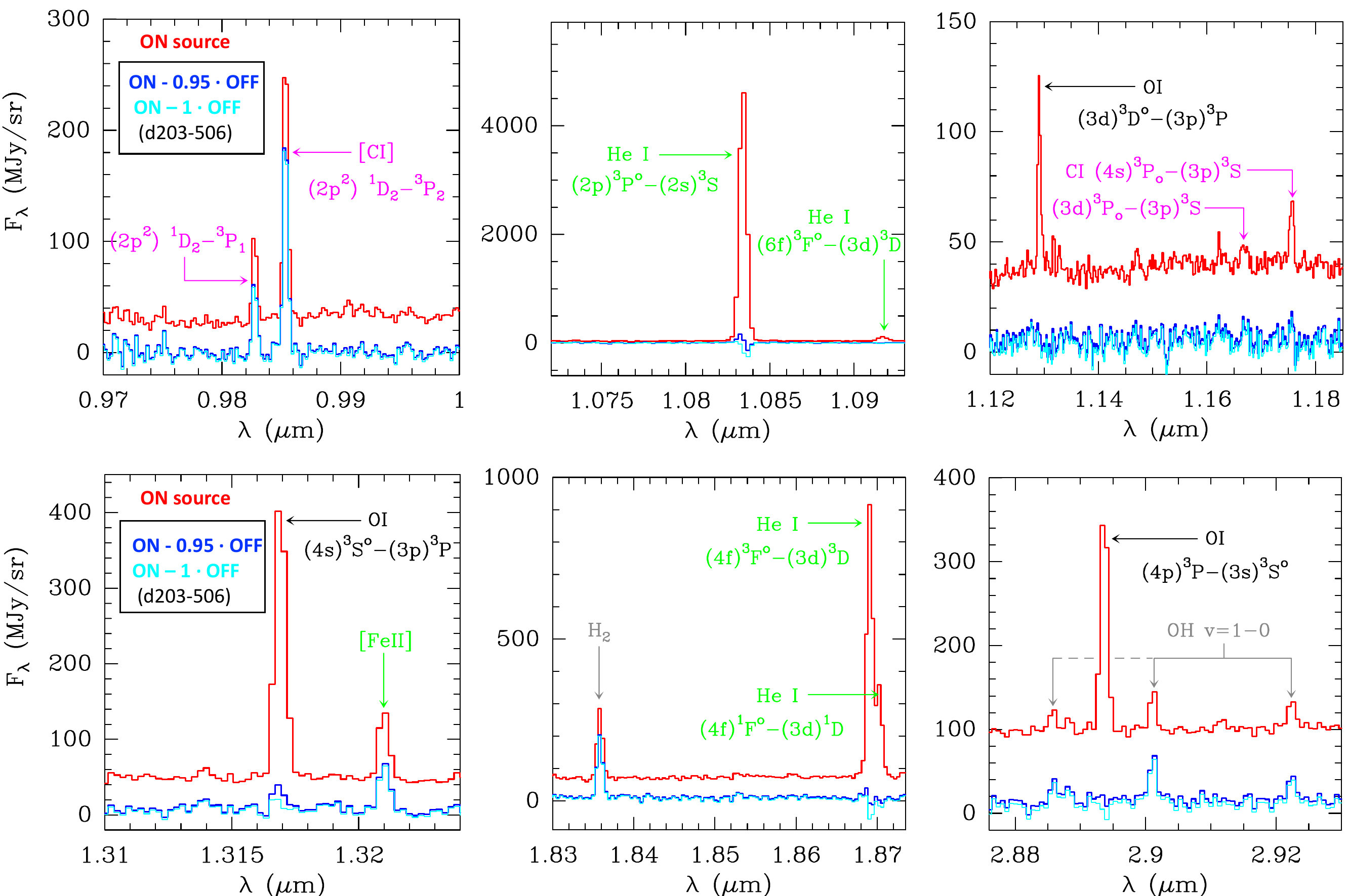} 
\caption{Selected JWST/NIRSpec spectra toward d203-506. The red spectrum corresponds
to the ON source measurement, which includes emission from the
background \HII~region and the Bar PDR. The intrinsic 
spectrum of \mbox{d203-506} is the {ON\,--\,$f$\,$\cdot$\,OFF} measurement
({with \mbox{$f$\,=\,0.95} in blue and \mbox{$f$\,=\,1} in cyan}).
The scaling factor, $f,$ is determined from the nebular \HeI~recombination lines (see the main text).  These spectra show
NIR  carbon lines (permitted and forbidden), [\FeII], and vibrationally excited H$_2$ and OH lines \citep{Zannese24}, among other species
\citep[see][]{Berne24}. However, NIR \OI~fluorescent lines are not detected in the \mbox{ON--OFF} spectrum. These FUV-pumped \OI~fluorescent lines appear in the ON spectrum and arise from predominantly atomic gas close behind the ionization front of the Bar
\citep[the $\Delta$ feature;][]{Haworth23,Peeters24}.}
\label{Fig:app-oi_spectra}
\end{figure*}

\section{Extraction of NIR spectra, background line emission, and FUV-pumping photons}
\label{app-spectrum extraction}

We first extracted the NIRSpec spectrum  toward \mbox{d203-506} \mbox{(ON source measurement)}.
As in our previous studies, we extracted the  ON  spectrum from an elliptical aperture centered on 
\mbox{$\alpha$(2000)\,=\,5$^{\rm h}$35$^{\rm m}$20.357$^{\rm s}$},
 \mbox{$\delta$(2000)\,=\,$-$5$^{\rm \circ}$25$'$05.81$''$} with dimension length 
 $l$\,=\,0.52$''$, height $h$\,=\,0.38$''$, and a PA=33$^o$  East of North \citep{Berne24}.
{This aperture includes both the \mbox{inner}, dense disk detectable in millimeter dust continuum emission \mbox{\citep{Berne24}}
and also the extended, optically thinner, outer disk and photo-evaporative wind.} 
Among other atomic ion lines, the ON spectrum contains a plethora of bright \HeI~lines.
 {These permitted NIR lines arise from very high-energy
levels, up to $\sim$24\,eV, thus they cannot be FUV-pumped lines of neutral helium.}
Instead, they are produced by  EUV photoionization of He atoms,  \mbox{IP(He)\,=\,24.6\,eV}, followed by radiative recombination cascades of He$^+$.
 {These \mbox{\HeI~recombination} lines are \mbox{nebular} lines \citep[][]{Rogers24}}  from the background   \HII~region \citep{Peeters24}. This implies
 that {toward the ON aperture}, \mbox{d203-506} is not 
 opaque to the background NIR line emission. 
In order to obtain the intrinsic  NIR spectrum of \mbox{d203-506}, one  needs to correct for any background \mbox{\HII~region} and Bar PDR line emission. As in our previous studies, we extracted a reference spectrum
very close to the disk and providing the \mbox{OFF-target} correction.  The OFF spectrum was obtained in a circular aperture of radius $r$\,=\,0.365$''$ centered  \mbox{$\alpha$(2000)\,=\,5$^{\rm h}$35$^{\rm m}$20.37$^{\rm s}$},
 \mbox{$\delta$(2000)\,=\,$-$5$^{\rm \circ}$25$'$04.97$''$}, that is, 
 at \mbox{(0.20$''$,\,0.84$''$)} from the disk  \citep[see][]{Berne24}.
 The adopted intrinsic spectrum of \mbox{d203-506} \mbox{(Fig.~\ref{fig:d203-506_spectra})} is the  {\mbox{ON\,--\,$f$$\cdot$\,OFF} spectrum, where $f$ is the scaling factor needed to have no \HeI~recombination line emission from \mbox{d203-506} itself. We find 
 \mbox{$f\simeq 0.95-1$}, similar to \cite{Rogers24} for other disks.}
 {This range provides a $\sim$5\,\% uncertainty in the line intensities of the NIR carbon lines,} which we derived  after baseline subtraction and Gaussian line fitting
 of the {\mbox{ON\,--\,$f$$\cdot$\,OFF} spectrum}.

The ON source spectrum also  shows several \OI~\mbox{permitted} lines: 
 \mbox{(3$d$)\,$^3$D$^{\rm o}$\,--\,(3$p$)\,$^3$P} at 1.129\,$\mu$m, 
 \mbox{(4$s$)\,$^3$S$^{\rm o}$\,--\,(3$p$)\,$^3$P} at 1.317\,\,$\mu$m, and
  \mbox{(4$p$)\,$^3$P\,--\,(3$s$)\,$^3$S$^{\rm o}$} at 2.894\,$\mu$m
(see \mbox{Fig.~\ref{Fig:app-oi_spectra}}), that are not seen in the spectrum of \mbox{d203-506} (\mbox{ON--OFF}). 
The above lines arise from the background, from the  predominantly neutral gas 
 (\mbox{$x_{\rm H^0}$\,$>$\,$x_{\rm H^+}$}) that borders close behind the
 ionization front of the Bar PDR \mbox{\citep{Walmsley00,Peeters24}}. These NIR \OI~lines are not O$^+$ recombination lines --consistent with the slightly higher IP
 of oxygen (13.618\,eV) compared to hydrogen (13.598\,eV)--, as
 there are no EUV photons capable   of ionizing neutral oxygen atoms inside PDRs. 
The excitation of the NIR \OI~lines in the Bar PDR is attributed to FUV-pumping followed by fluorescence \citep{Walmsley00,Lucy02,Henney21,Peeters24}. As for neutral carbon, the ground-state of neutral oxygen is a triplet $^3$P state, so that
 FUV excitations occur for high triplet 
 states. \mbox{However},
 the main pumping lines of the NIR \OI~lines {at 1.129, 1.317, and  2.894\,$\mu$m} lie at short FUV wavelengths: respectively at  1027\,$\AA$ (12.1\,eV; {around Lyman-$\beta$}), 1040\,$\AA$ (11.9\,eV),
 and multiple lines between 918 and 979\,$\AA$ (13.5--12.7\,eV). 
In a neutral PDR, these high-energy FUV photons 
(significantly more energetic than the \mbox{7.5-9.6\,eV} photons needed to  pump  C$^0$)
 are only available close to the ionization front, where the gas is mostly atomic.

 The lack of remarkable NIR
\OI~emission in  \mbox{d203-506} {(other than a  faint \OI~1.317\,$\mu$m emission
in the \mbox{ON--\,0.95\,$\cdot$\,OFF} spectrum; Fig.~\ref{Fig:app-oi_spectra})} is not due to a reduced gas-phase elemental abundance of oxygen. Instead, it is caused by a reduced flux of FUV photons with energies capable of pumping the \OI~levels that are permeating the disk.
This deficiency leads to very faint NIR \OI~fluorescent emission, likely confined to a very thin layer on the outermost surface of the wind.
Hence, we conclude that \mbox{d203-506} is not pervaded by a high flux of  FUV photons with energies above  $\sim$12\,eV: both FUV continuum photons and Lyman--$\beta$ photons at 1026\,\AA~(\mbox{accidental resonance} with the {main FUV-pumping line
of the NIR \OI~\,1.317\,$\mu$m}  emission  line
 in  a Bowen-fluorescence mechanism). These photons are  absorbed close to the 
  wind surface layers. Indeed, at high \mbox{$n_{\rm H}$/$G_0$} ratios ($\gtrsim$\,500\,cm$^{-3}$) and relatively low effective dust FUV \mbox{absorption} cross sections,  
these photons   are   removed by gas absorption: C$^0$ photoionization continuum (for \mbox{$\lambda$\,$<$1101\,\AA}; \mbox{$E$\,$>$\,11.26\,eV}), \mbox{Lyman~\HI}, and \mbox{Lyman-Werner} H$_2$ lines (see  \mbox{Fig.~\ref{Fig:FUV_field_disk}}), which develop broad line-absorption wings as gas column densities sharply  increase \mbox{\citep[e.g.,][]{Goico07}}. In \mbox{Appendix~\ref{app-new-FUV-pumping-model}}, we 
explicitly model these processes for the
physical conditions in \mbox{d203-506}.
In addition, a small fraction of  $\gtrsim$12\,eV photons might be absorbed by any dust and neutral gas component situated between \mbox{d203-506} and the massive stars \mbox{$\theta^1$ Ori C} and \mbox{$\theta^2$ Ori A}  \citep[e.g.,][]{vanderWerf13,ODell20,Pabst20}.

The above reasoning is also supported by the lack of fluorescent \OI~\mbox{(3$p$)\,$^3$P\,--\,(3$s$)\,$^3$S$^{\rm o}$}  line emission at \,0.845\,$\mu$m   \citep[VLT/MUSE   
images, see][]{Haworth23}, where the upper energy level of the \mbox{\OI~\,0.845\,$\mu$m} transition is the lower-energy level of the NIR \OI~\,1.129\,$\mu$m and 1.317\,$\mu$m
lines targeted by \mbox{NIRSpec} 
        \citep[i.e., the \mbox{\OI~\,0.845\,$\mu$m} line 
is pumped by the same $\gtrsim$12\,eV  FUV-photons than the \OI~lines targeted by NIRSpec,][]{Bowen28,Walmsley00,Henney21}.
 Instead,  the disk is seen in silhouette   against the background 
\mbox{\OI~\,0.845\,$\mu$m} line  emission from the rim of the Bar PDR \citep[the \mbox{$\Delta$-feature},][]{Haworth23}.
Hence, most of the NIR \OI~fluorescent emission in the ON
spectrum arises from this background feature \citep[Fig.~8 of ][]{Peeters24}.

A reduced number of $\gtrsim$\,12\,eV photons  also implies that CO (X$^1$$\Sigma^+$) photodissociation
will mainly produce neutral carbon in the ground $^3$P state 
\citep[\mbox{dissociation} energy of 11.1\,eV, e.g.,][]{vD88}.
Indeed,  experiments show that FUV photons with energies $\geq$12.4\,eV are needed to produce neutral carbon atoms directly in the first electronically excited $^1$D state  \citep{Guan21}. Thus, this mechanism will be less dominant in  \mbox{d203-506}.

\section{Recombination theory applied to the observed NIR carbon lines}
\label{app-recomb-theory}

\begin{figure}[t]
\centering   
\includegraphics[scale=0.7, angle=0]{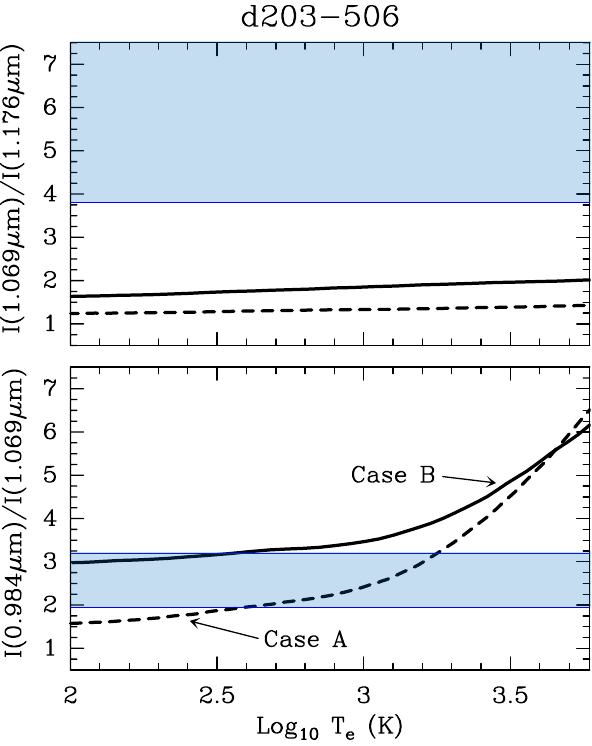}
\caption{NIR carbon line intensity ratios predicted by the recombination theory of  \cite{Escalante90}. Curves represent the expected ratios  under Case B (continuous)
and Case A (dashed) conditions, taken from \cite{Walmsley00}. 
The blue-shaded area shows the line ratios in \mbox{d203-506} that are compatible with the observations. See the main text.}
\label{Fig:ir_carbon}
\end{figure}

\mbox{Figure~\ref{Fig:ir_carbon}} is analogous to Fig.~14 of \cite{Walmsley00}.
It shows the \mbox{$I$(0.984\,$\mu$m)/$I$(1.069\,$\mu$m)} (bottom)
and \mbox{$I$(1.069\,$\mu$m)/$I$(1.176\,$\mu$m)} (top)
 line
intensity ratios predicted by
the recombination theory \citep[\mbox{C$^+$\,+\,e$^-$\,$\rightarrow$\,C$^*$\,+\,line cascade};][]{Escalante90,Escalante91} in two limiting cases:
Case~A (\mbox{optically thin} gas in all transitions) and the more realistic Case~B
(\mbox{optically thick} gas to photons produced by permitted transitions, and direct radiative recombinations  to the ground state).
Here, the $I$(0.984\,$\mu$m)  refers to the intensities
of the observed \mbox{0.9827\,$+$\,0.9853\,$\mu$m} forbidden lines, and $I$(1.069\,$\mu$m) and
$I$(1.176\,$\mu$m) are the sum of all permitted line intensities in these multiplets
(see Table~\ref{Table_intensities}).

In the recombination theory,
the \mbox{$I$(1.069\,$\mu$m)/$I$(1.176\,$\mu$m)}  ratio depends mainly on the opacity
of the FUV-resonant transitions. However, the observed \mbox{$I$(1.069\,$\mu$m)/$I$(1.176\,$\mu$m)}  ratio in \mbox{d203-506}  
(9.8\,$\pm$\,6.0; blue-shaded areas) is significantly above the ratio predicted by this theory. 
Furthermore, the \mbox{$I$(0.984\,$\mu$m)/$I$(1.069\,$\mu$m)} line intensity ratio
 depends on the electron temperature.  
Without any extinction correction, the observed ratio only fits the predictions
of Case A, which is less realistic. 
 The upper observational value
of the \mbox{$I$(0.984\,$\mu$m)/$I$(1.069\,$\mu$m)}  ratio in \mbox{Fig.~\ref{Fig:ir_carbon}}
 includes a foreground extinction correction of 1.5\,mag 
\citep[][]{Peeters24} adopting the $R_V$\,=\,5.5  extinction law of \cite{Gordon23}.
In any case,  the absolute intensities of the NIR C$^0$ lines observed  in \mbox{d203-506}  are stronger (see the next section and Table~\ref{Table_intensities}) than those predicted
by C$^+$ recombination theory {\citep[e.g.,][]{Escalante91}}.
This comparison implies that the involved C$^0$ levels should be predominantly populated by means other than recombination.

\clearpage

\section{Inclusion  of FUV-pumping for C$^0$ levels and C$^+$\,+\,e$^-$ recombination followed by radiative cascades in the Meudon PDR code}
\label{app-new-FUV-pumping-model}

\begin{figure}[t]
\centering   
\includegraphics[scale=0.48, angle=0]{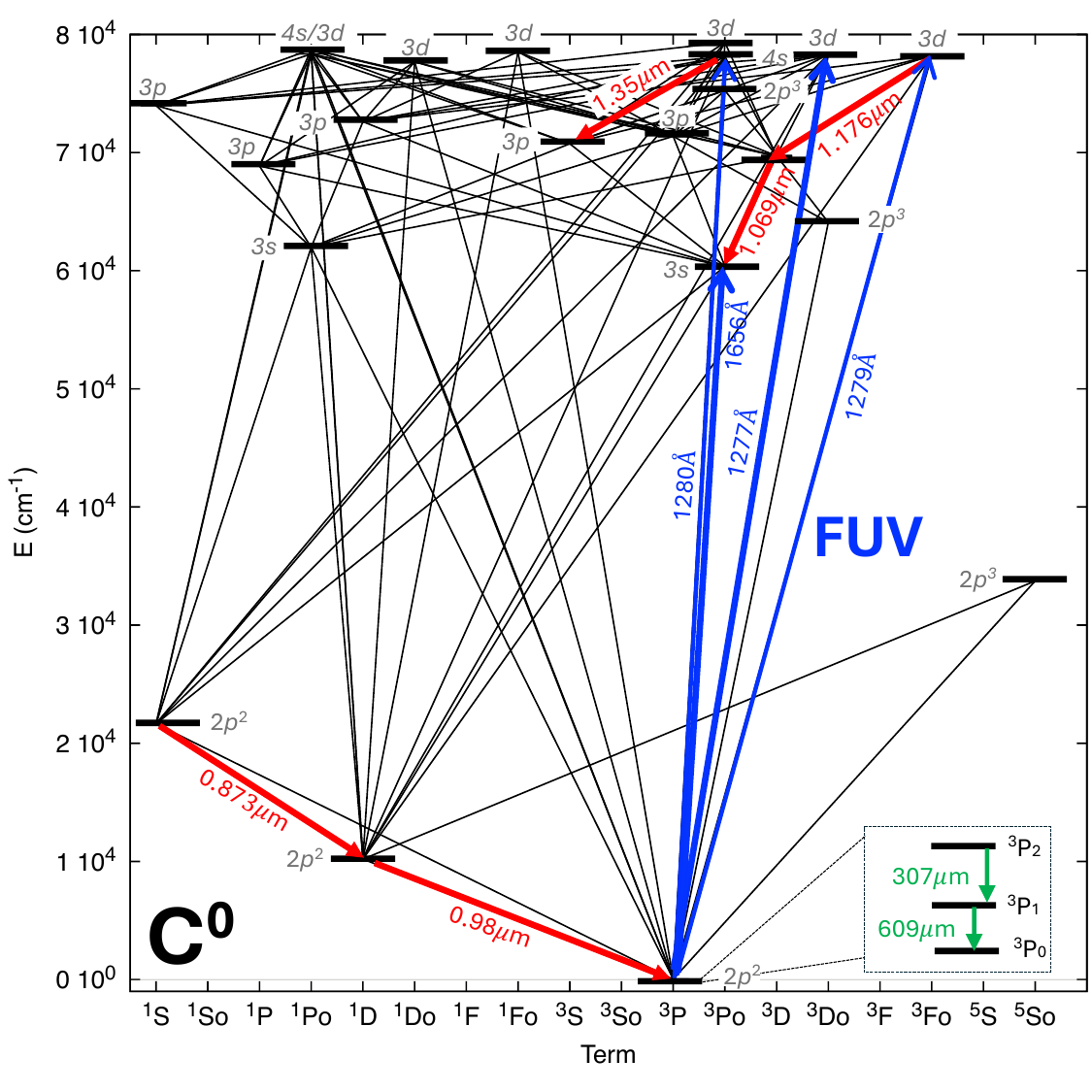}
\caption{Reduced Gotrian diagram of C$^0$ showing up to the electronic  levels associated with the
NIR lines discussed in this study. The figure also includes the main \mbox{FUV-pumping} lines
(in blue) of the detected NIR C$^0$ lines, as well as the submm fine-structure lines
within the ground state detectable with ALMA
 (in green; energy level splitting is exaggerated).}
\label{Fig:energy_levels_all}
\end{figure}

\begin{figure*}[t]
\centering   
\includegraphics[scale=0.68, angle=0]{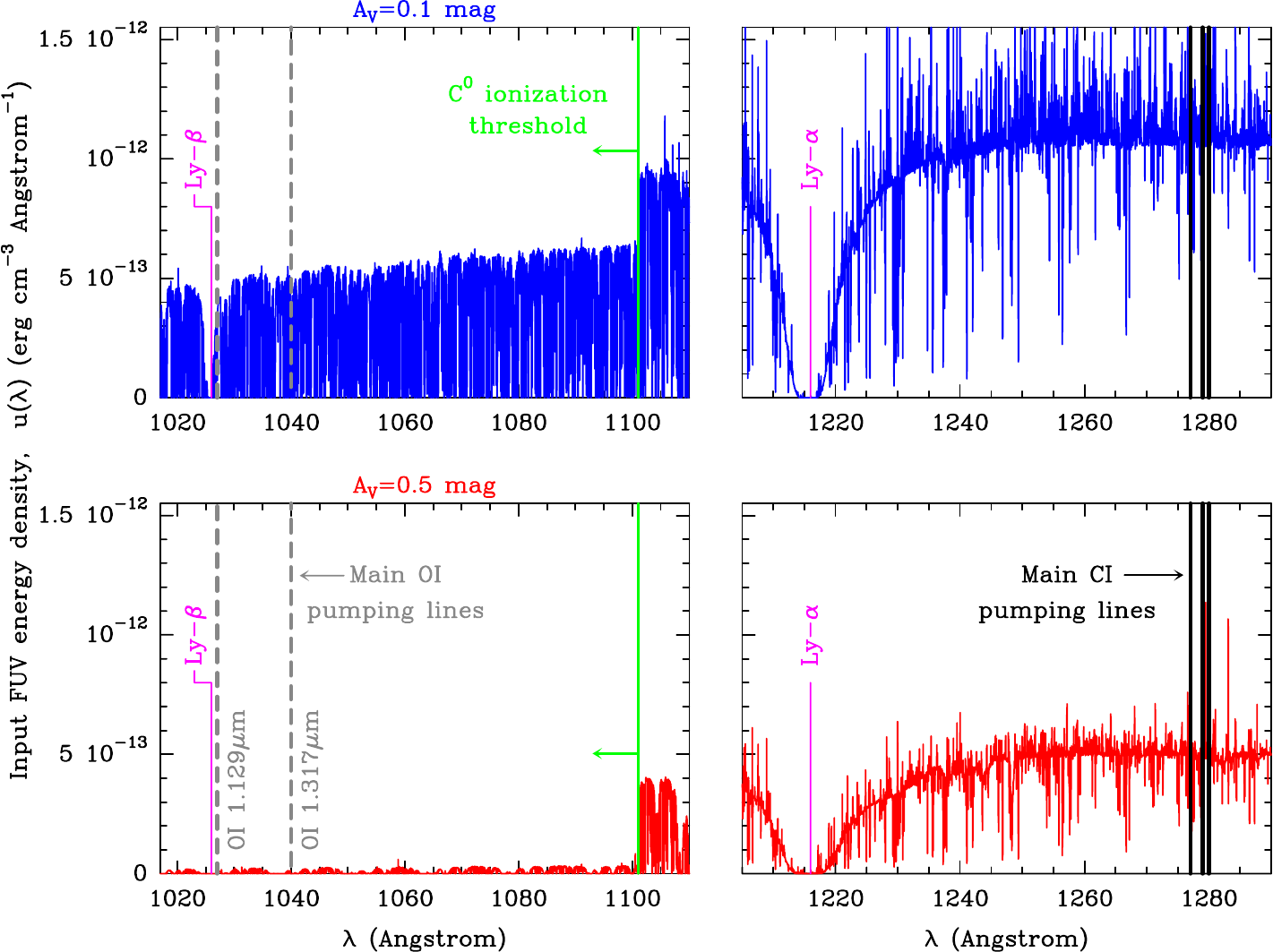}
\caption{External FUV radiation field (in energy density) propagating inside the
{photoevaporating wind and outer disk}  (reference model of \mbox{d203-506} with \mbox{$G_0$\,=\,2\,$\times$\,10$^4$} and \mbox{$n_{\rm H}$\,=\,10$^7$\,cm$^{-3}$}). 
The upper panels (blue spectra) are for a slab position at $A_V$\,=\,0.1\,mag (close to the PDR surface), whereas the lower panels (red spectra) are for a position deeper inside the PDR, at $A_V$\,=\,0.5\,mag.
The disk-depth-dependent FUV field results from a FUV radiative transfer calculation that includes gas absorption, H$_2$ fluorescent emission, and grain absorption and scattering. The resulting FUV continuum and line emission are modulated by strong  \HI~absorption lines (Ly-$\alpha$ and Ly-$\beta$ lines are indicated in this figure) and by a forest of H$_2$ absorption and fluorescent emission lines.
The black vertical  lines show the wavelength position of the main C$^0$~pumping transitions
at 1277, 1279, and 1280\,\AA~($\sim$9.6\,eV).  The vertical dashed gray lines 
show the wavelength position of the main O$^0$~pumping transitions
at 1027 and 1040\,\AA~($\sim$12\,eV). These photons are quickly absorbed by small column densities of gas,
mainly via C$^0$ photoionization below 1101\,\AA\  for $A_V$\,$<$\,1\,mag (\mbox{Fig.~\ref{Fig:Absorption_coeff}}).
Lower-energy FUV photons propagate deeper inside the disk and are attenuated by dust grains. 
}
\label{Fig:FUV_field_disk}
\end{figure*}

Prompted by the impossibility of explaining the observed NIR carbon lines within the recombination theory, and drawing an analogy with the \mbox{FUV-pumping} mechanism of NIR \OI~\,lines at 1.129, 1.317, and 2.894\,$\mu$m  seen
toward the ionization fronts of \mbox{interstellar} PDRs  \citep[e.g.,][]{Bowen28,Walmsley00,Lucy02,Henney21},
 we incorporated the FUV radiative excitation of C$^0$    in the Meudon PDR code 
 \mbox{\citep{LePetit06}}. We specifically included 401 levels of C$^0$ (electronic states and their
fine-structure levels), with electronic configurations up to 29$d$ (at $\sim$11.25\,eV)
and involving a total of {1572} radiative transitions. 
Figure~\ref{Fig:energy_levels_all} shows a reduced Gotrian diagram of C$^0$, showing up to
the electronic levels associated with the NIR lines detected with NIRSpec.
We took the main spectroscopic parameters from the NIST data 
base\footnote{\href{URL}{https:$//$physics.nist.gov$/$PhysRefData$/$ASD$/$levels$\_$form.html}}.
The  external FUV  field ($\lambda$\,$>$\,912\,\AA) penetrating \mbox{d203-506} is self-consistently computed
as in \cite{Goico07}. The attenuation of  FUV continuum and line  photons  impinging in \mbox{d203-506} includes absorption  and anisotropic scattering by dust grains as well as 
gas FUV-line absorption and photoionization {continuum}. The disk-position and {$\lambda$}-dependent FUV-field couples to C$^0$  by the resonant absorption of FUV photons from the $^3$P ground-state  to high-energy electronic triplet states
through several pumping transitions: 1656\,\AA, 1277\,\AA, 1279\,\AA, 1280\,\AA, etc.
(see \mbox{Fig.~\ref{Fig:FUV_field_disk}}). Subsequent
visible and NIR fluorescent line emission cascades populate the  lower-energy C$^0$ levels.
The brightest NIR carbon lines predicted by the model are precisely the ones detected by 
JWST/NIRSpec and VLT/MUSE: the forbidden lines at $\sim$0.984 and 0.8729\,$\mu$m, as well as the higher-energy permitted lines at  $\sim$1.069, $\sim$1.176 , and $\sim$1.355\,$\mu$m, in decreasing order of intensity.
The model also predicts that other NIR \CI~permitted line multiplets  should be detectable
(e.g., at $\sim$0.910, $\sim$0.966, and $\sim$1.455\,$\mu$m). However, the NIRSpec observation 
of \mbox{d203-506} does not cover these specific wavelength ranges.

In addition, we included a reduced description of \mbox{C$^+$\,+\,e$^-$} radiative recombination, including cascades toward the lowest 254 levels of C$^0$ 
(with core electrons in one of the ground-state $^2$P$_{3/2}$ or $^2$P$_{1/2}$
fine-structure levels of C$^+$). We adopted the recombination rates
from \cite{Badnell06}.
However, for the physical conditions of the NIR-emitting gas layers of \mbox{d203-506} 
($G_0$\,of few 10$^4$ and \mbox{$n_{\rm H}$\,$\simeq$\,10$^7$\,cm$^{-3}$}), we find that
\mbox{C$^+$\,+\,e$^-$} recombinations  only increase the intensity of the observed
NIR carbon lines by $\lesssim$\,5\,\%. Therefore, in \mbox{d203-506}, this is not the dominant excitation process
of these NIR lines.

Our excitation model also incorporates inelastic collisions of C$^0$ with e$^-$, from the 
ground-state  $^3$P to the first and second excited C$^0$ electronic levels,  $^1$D$_2$ and
$^1$S$_0$ \citep[rate coefficients from][]{Pequignot76,Mendoza83}. Nevertheless, we find that collisional excitation to these high-energy
levels plays a negligible role compared to \mbox{FUV-pumping}. 
Within the ground-state $^3$P$_J$ fine-structure levels, the model also includes inelastic collisions with e$^-$ \citep{Johnson87}, with H \citep{Abrahamsson07}, with He \citep{Staemmler91},
and with \mbox{$o$-/$p$-H$_2$}  \citep[][and extrapolated to high temperatures as in Goicoechea et al. in prep.]{Plomp23}. However,  at the high densities in 
disks 
($n_{\rm H}$\,$\gg$\,$n_{\rm cr,\,[CI]609}$\,$\simeq$\,10$^3$\,cm$^{-3}$), the [\CI]609\,$\mu$m fine-structure line is thermalized ($T_{\rm ex}$\,=\,$T_{\rm k}$), and its
excitation is independent on
 the collisional rate coefficients. [\CI]\,370 and 609\,$\mu$m line  intensities are not affected by FUV-pumping either.

Our PDR models including FUV-pumping and fluorescent de-excitation
match the intensity level of the observed NIR carbon lines. Without this radiative excitation mechanism, the predicted NIR line intensities would be up to two orders of magnitude lower
than observed (Table~\ref{Table_intensities}). FUV-pumping also produces higher  values of the \mbox{$I$(1.069\,$\mu$m)/$I$(1.176\,$\mu$m)} ratio than those predicted by recombination theory.  The reference model predicts that the observed NIR  
and  [\CI]\,609\,$\mu$m lines are optically thin, with \mbox{$\tau$\,$\simeq$\,10$^{-6}$}
and \mbox{$\tau$\,$\simeq$\,10$^{-1}$}, respectively.
In contrast, the main \CI~\,\mbox{FUV-pumping} lines at
1656\,\AA, 1277\,\AA, 1279\,\AA, and 1280\,\AA~have very high opacities, ranging from \mbox{$\tau$\,$\simeq$\,200} to $\simeq$\,5000. The FUV continuum is optically thick, 
\mbox{$\tau$\,$\simeq$\,30-40} (see \mbox{Fig.~\ref{Fig:FUV_field_disk}}).

In the Orion Bar PDR, \cite{Peeters24}  found 
high \mbox{$I$(0.984\,$\mu$m)/$I$(1.069\,$\mu$m)} line ratios and 
 \mbox{$I$(1.069\,$\mu$m)/$I$(1.176\,$\mu$m)}  line ratios  that are inconsistent with recombination theory too (their Fig.~G.1). This suggests that recombination alone may not drive the excitation of NIR carbon lines in strongly irradiated interstellar PDRs either. 
Adding the role of FUV-pumping will likely relax  the hot temperatures (several thousand K)
and densities (\mbox{$\sim$\,10$^8$\,cm$^{-3}$}) determined by \cite{Peeters24}. 
That is, there will be no need to invoke the presence of very high-pressure small clumps, which are otherwise not seen in the JWST images.

\section{FUV attenuation by gas lines and dust}
\label{app-gas_absorption}

\begin{figure}[h]
\centering   
\includegraphics[scale=0.55, angle=0]{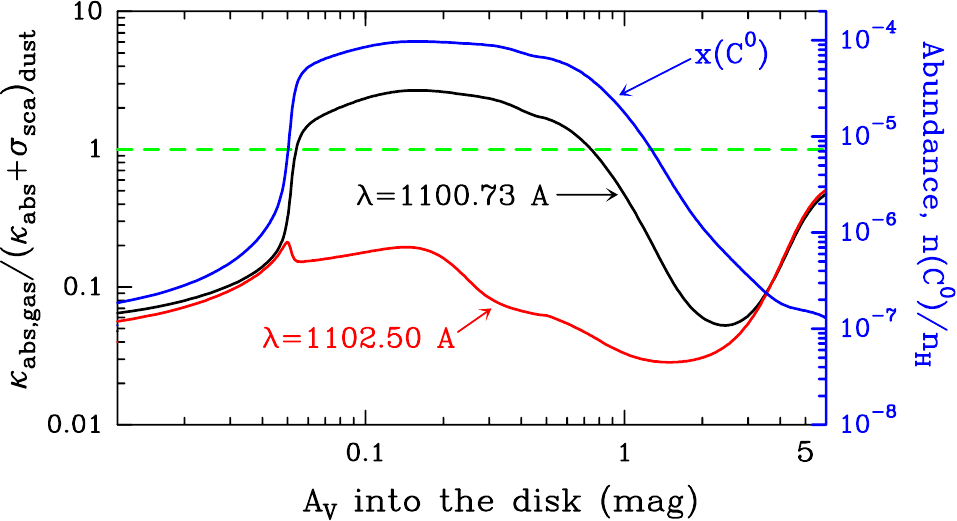}
\caption{Evolution of the ratio between the wavelength-dependent FUV gas absorption coefficient and the dust extinction coefficient
(absorption plus scattering) as a function of $A_V$ for two representative continuum  wavelengths: one
below the C$^0$ photoionization threshold (black curve) and one above (red curve). 
For  positions where the black curve  is greater than 1, C$^0$ photoionization
dominates the absorption of FUV photons with $\lambda$\,$<$\,1101\,\AA. This process 
 reduces  the amount of FUV photons that pump the NIR  oxygen lines close to  the PDR surface.}
\label{Fig:Absorption_coeff}
\end{figure}

In this section we provide evidence that in irradiated protoplanetary disks,  FUV photons at the pumping transitions of the \OI~\mbox{fluorescent} lines (photons with 
$E\gtrsim $\,12\,eV) can be more quickly removed  than photons
pumping neutral carbon ($E\lesssim $\,9.6\,eV).
Compared to interstellar PDRs, where small dust grains dominate the absorption of FUV photons, gas absorption becomes very important in the upper layers of irradiated protoplanetary disks. This is due to the reduced {effective} FUV dust absorption cross section, caused by larger grain sizes and  possibly higher gas-to-dust mass ratios {than in the ISM}.
 We determine that, close to the PDR surface of \mbox{d203-506}
(from  \mbox{$A_V$\,$\simeq$\,0.05} to $\simeq$\,0.8\,mag), C$^0$ photoionization  dominates the absorption of
FUV photons with $\lambda$\,$<$\,1101\,\AA~($>$11.26\,eV). For longer photon wavelengths and for PDR depths higher than  
\mbox{$A_V$\,$\simeq$\,0.8}\,mag, dust absorption dominates.
To be more quantitative, \mbox{Fig.~\ref{Fig:Absorption_coeff}} shows the evolution of the ratio of the
FUV gas absorption coefficient over the dust extinction coefficient as a function of $A_V$ for two representative
continuum wavelengths: 1100.73\,\AA~(below the C$^0$ ionization threshold; in black)
and 1102.50\,\AA~(above this threshold, i.e., not producing C$^0$ ionization; in red). 
The black curve shows that this ratio reaches nearly a factor of 3 (that is, C$^0$ photoionization
dominates over FUV dust absorption), thus producing enhanced absorption of 
$<$\,1101\,\AA~photons compared to photons at longer wavelengths. This is exemplified by the sharp step, delineated by the green vertical line, in the FUV spectrum of Fig.~\ref{Fig:FUV_field_disk}. This plot shows that FUV photons at the 
\mbox{pumping} transitions of neutral oxygen atoms ($\sim$1000\,\AA) are scarcer than those at the pumping transitions of neutral
carbon ($\sim$1300\,\AA).

\section{Variations in \mbox{$I$([\CI]\,609\,$\mu$m)} and NIR carbon lines with FUV flux and gas density}
\label{app-atomic_carbon}

\begin{figure}[h]
\centering   
\includegraphics[scale=0.45, angle=0]{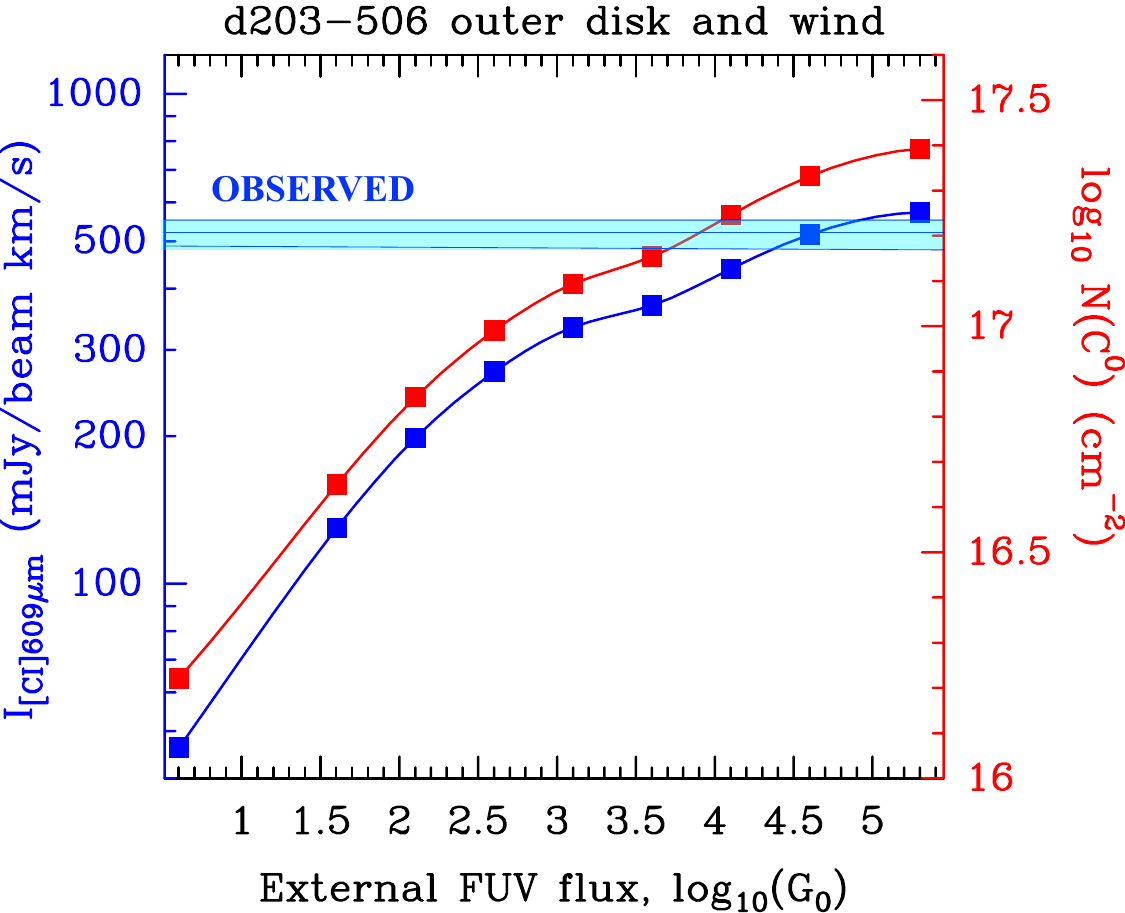}
\caption{Grid of constant-density PDR models
for varying external FUV radiation
fields but fixed $n_{\rm H}$\,=\,10$^7$\,cm$^{-3}$ and $x_{\rm C}$\,=\,1.4\,$\times$\,10$^{-4}$ values.
 The red markers  show the predicted
column density of C$^0$. The blue markers  show
the predicted [\CI]\,609\,$\mu$m line intensity,  integrating from
$A_V$\,=\,0 to 10 mag into the wind and disk system.
The horizontal lines mark the observed line intensity ($\pm$\,$\sigma$) in \mbox{d203-506}.}
\label{Fig:app-pdr-carbon}
\end{figure}

{In this study, we assume that at the physical scales probed by our observations, the density distribution is smooth across the outer disk and inner wind. Therefore, as a first approximation,  we adopt constant density models.}
In PDR environments, the C$^0$ column density, and thus the intensity of the \mbox{[\CI]\,609\,$\mu$m}
line, are not strong
functions of the external FUV flux or gas density.
In general, $N$(C$^0$) somewhat decreases with increasing $n_{\rm H}$ and slightly
increases with increasing $G_0$.
For the irradiation conditions {in the outer layers of} \mbox{d203-506}, $G_0$ of a few 10$^4$, and gas densities varying between $n_{\rm H}$\,=\,10$^6$\,cm$^{-3}$ and  10$^8$\,cm$^{-3}$, the predicted line intensity \mbox{$I$([\CI]\,609\,$\mu$m)} only changes by a factor of 3.
Adopting $n_{\rm H}$\,=\,10$^{7}$\,cm$^{-3}$  in the outer disk  and inner wind, 
\mbox{$I$([\CI]\,609\,$\mu$m)} only increases by a factor of $\lesssim$\,2
when the external FUV flux changes from $G_0$\,=\,10$^3$ and 10$^5$
(see \mbox{Fig.~\ref{Fig:app-pdr-carbon}}). Therefore,  
 with a reasonable knowledge of $G_0$ and $n_{\rm H}$, the  \mbox{$I$([\CI]\,609\,$\mu$m)} line intensity can be used to accurately determine the column density of C$^0$ and, through modeling,  the  \mbox{(total)} gas-phase \mbox{carbon} abundance, 
\mbox{$x_{\rm C}$\,=\,$x_{\rm C^+}$\,+\,$x_{\rm C^0}$\,+\,$x_{\rm CO}$\,+\,...\,}, in the 
[\CI]\,609\,$\mu$m-emitting  layers.

\begin{figure}[h]
\centering   
\includegraphics[scale=0.65, angle=0]{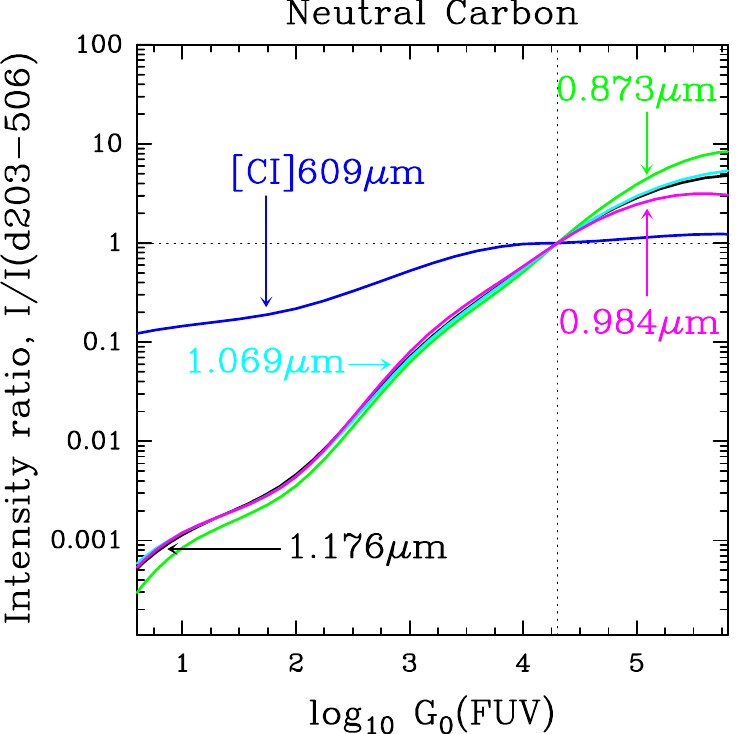}
\caption{Grid of constant-density PDR models
for varying external FUV radiation
fields but fixed $n_{\rm H}$\,=\,10$^7$\,cm$^{-3}$ and $x_{\rm C}$\,=\,1.4\,$\times$\,10$^{-4}$. Colored  curves show the predicted line intensities of the neutral carbon lines
discussed in the text relative to the line intensity reference model 
($G_0$\,=\,2\,$\times$\,10$^4$). While the NIR carbon line intensities increase with $G_0$, the
[\CI]\,609\,$\mu$m line intensity is much less dependent on $G_0$.}
\label{Fig:app-fuv-pumping-G0}
\end{figure}

In our models, however, the NIR carbon lines follow a different emission trend as they are mostly sensitive to the flux of FUV-pumping photons (i.e., to $G_0$; \mbox{see Fig.~\ref{Fig:app-fuv-pumping-G0}}). 
This is a consequence of the negligible collisional (de)excitation of the NIR C$^0$ lines
($n_{\rm H}$\,$\ll$\,$n_{\rm cr}$) and the attenuation of the \mbox{FUV-pumping photons}
at high gas densities. 
At high $G_0$, this makes the column density
of C$^0$ that is sensitive to FUV-pumping, $N$(C$^{0}_{\rm pump}$), roughly proportional to  $G_0$/$n_{\rm H}$, and the intensity of the NIR C$^0$~fluorescent lines approximately  proportional to 
\mbox{$n_{\rm H}$\,$N$(C$^{0}_{\rm pump}$)\,$\simeq$\,$G_0$}.
This conclusion applies to both the  high-energy \CI~permitted lines
and also the [\CI]~\mbox{forbidden} lines at 0.8729\,$\mu$m and $\sim$0.984\,$\mu$m
\citep[we recall the later are not observed in the innermost regions, $<$\,1\,au, of \mbox{isolated} protoplanetary disks;][{but are bright in externally irradiated disks}]{
McClure19}.
Therefore, the NIR C$^0$ line intensities depend less on the gas density; they only change by $\sim$\,30\,\%~ if the gas density changes by an order of magnitude compared to the reference model.
For the same reason, a reduction of an order of magnitude  in the  total gas-phase carbon abundance,
$x_{\rm C}$, only changes the intensity of the NIR carbon lines by a factor of $\lesssim$\,2.
Thus, we conclude that the extended emission from these NIR lines is a powerful probe of external FUV radiation fields impinging on protoplanetary disks in cluster environments.

\clearpage

\section{Little H$_2$O freeze-out in the outer disk and wind layers
traced by [\CI]\,609\,$\mu$m}
\label{app-snowline}

\begin{figure}[h]
\centering   
\includegraphics[scale=0.65, angle=0]{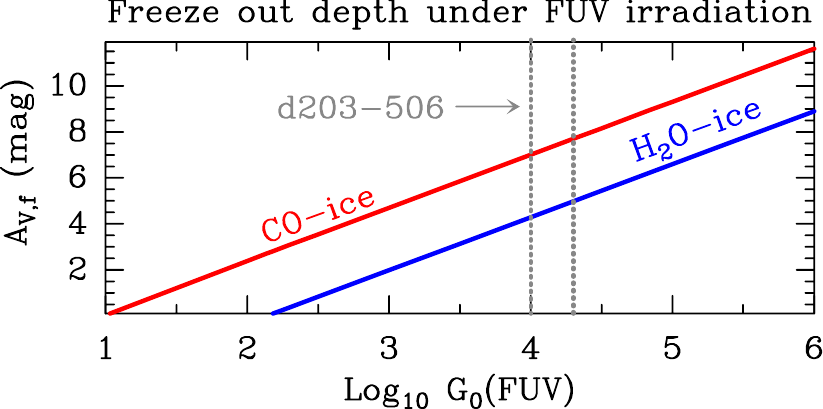}
\caption{Shift in the H$_2$O and {CO} freeze-out depth (simulating {the outer} disk)  with increasing external FUV radiation field. This analytical model
assumes $n_{\rm H}$\,=\,10$^7$\,cm$^{-3}$, as in d203-506. {The  vertical lines
show the $G_0$ values compatible with the observed NIR carbon line intensities.}} 
\label{Fig:snowline}
\end{figure}

Water ice is the first ice mantle to condense when dust temperatures cool below $T_{\rm d}$\,$\lesssim$\,130\,K \citep[depending on the assumed gas density, e.g.,][]{EvD21}. 
\mbox{However}, water ice \mbox{photo-desorption} and {thermal desorption}  replenishes the 
 gas-phase oxygen abundance and reduces the 
gas-phase $x_{\rm C}/x_{\rm O}$ ratio.
\mbox{Figure~\ref{Fig:snowline}} shows a simple analytical model of the predicted shift of the water freeze out depth ($A_{V,\,f}$) at which  most undepleted  (i.e.,~not in refractory material) gas-phase oxygen is incorporated into water ice mantles. This plot shows $A_{V,\,f}$, in mag of visual extinction  from the edge of the wind/disk system. We plot $A_{V,f}$ as a function of the external FUV photon flux, in $G_0$ units,
for a typical gas density in the outer disk and inner wind, $n_{\rm H}$\,=\,10$^7$\,cm$^{-3}$, and $T_{\rm k}$\,=\,$T_{\rm d}$. 
{This analytical determination of the H$_2$O freeze out depth implies} equating the rate at which undepleted gas-phase O atoms stick on grains, $R_{\rm gr, O}$ (first step in the grain surface formation of water ice) and the photo-desorption rate of water molecules from icy grain surfaces, 
$R_{\rm photodes,\,H_2O}$. These rates go
as $R_{\rm gr, O}\propto n_{\rm H}\,T_{\rm k}^{1/2}\,x_{\rm{O}}$
and $R_{\rm photodes,\,H_2O} \propto Y\,G_0\,\rm{exp}(-\it{b}\,A_V)$, respectively, where
$x_{\rm{O}}$ is the undepleted gas-phase abundance of O atoms, $Y$ is the number of desorbed molecules per incident FUV photon, and $b$ is a dust-related FUV field absorption factor
\citep[for the basic formalism and standard parameters see e.g.,][]{Hollenbach09}.
The water freeze out takes place very close to the irradiated disk surface
for  high values of the gas density over the FUV flux ratio, \mbox{$n_{\rm H}/G_0 \gtrsim 10^5$\,cm$^{-3}$}. However, for strong external FUV radiation fields (\mbox{$G_0 > 10^2$} in our representative example), the 
\mbox{$n_{\rm H}/G_0$} ratio decreases, and water ice only  becomes an abundant oxygen-reservoir deeper inside the disk. 
In this toy model of the outer disk layers of \mbox{d203-506}, the water freeze out depth
shifts by $\simeq$\,4.5\,mag {(tens of au)}
compared to isolated  disks, that is,  beyond
the [\CI]\,609\,$\mu$m emitting layers  (Fig.~\ref{fig:PDR_structure_203-506}). 
{The blue curve in \mbox{Fig.~\ref{Fig:snowline}} also shows the analytical estimation of the CO freeze-out depth using the rate at which gas-phase CO molecules stick on grains}.
Nearly all volatile oxygen and carbon in the
 disk zone traced by   [\CI]\,609\,$\mu$m and NIR C$^0$ line observations is in the gas-phase,
 and the gaseous elemental $x_{\rm C}/x_{\rm O}$ abundance ratio is low, close to the solar value
 in the case of Orion.
 
 Our more detailed  thermo-photochemical PDR models  in
 \mbox{Fig.~\ref{fig:PDR_structure_203-506}} include gas-phase and {simple gas-grain exchanges}  for O, OH, H$_2$O, O$_2$,  and CO.
These species adsorb on dust grains as temperatures drop, are photo-desorbed by FUV photons, desorb via cosmic-ray impacts, and thermally sublimate {\citep[see][for the inclusion of these processes in the Meudon PDR code]{Putaud19}.
Only for water ice formation we included the grain surface reactions
\mbox{s-H\,+\,s-O\,$\rightarrow$\,s-OH} and \mbox{s-H\,+\,s-OH $\rightarrow$\,s-H$_2$O}, where \mbox{s-} refers to the species in the solid \citep[see][]{Goicoechea21b}.} 
Thus, while the water ice abundance and  H$_2$O freeze-out depths are self-consistently
calculated, {more detailed surface reactions and models will be needed  to
accurately describe the growth and (photo)processing of carbon-bearing ices under strong
external FUV irradiation}.

\begin{figure}[ht]
\centering   
\includegraphics[scale=0.45, angle=0]{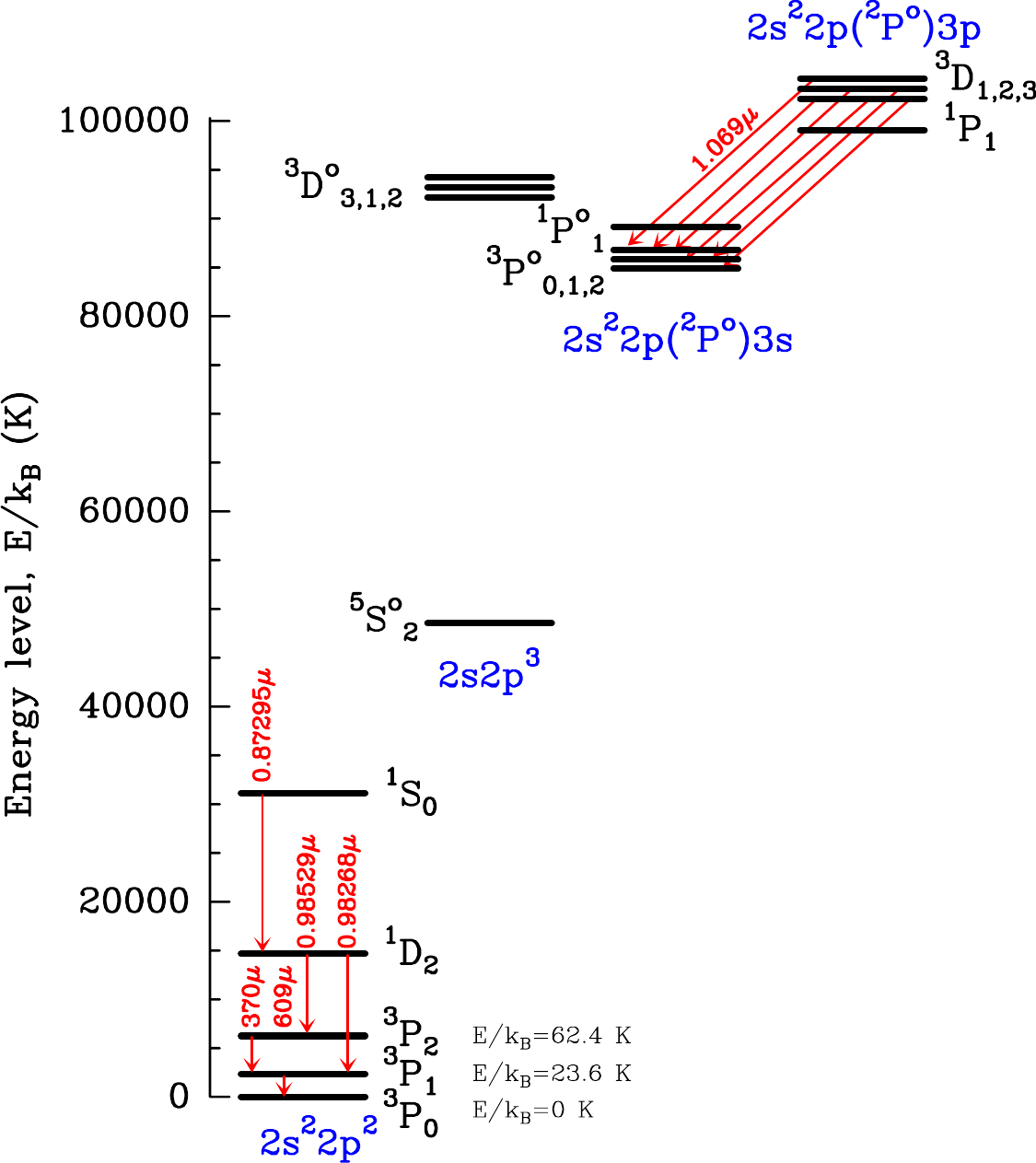}
\caption{Lowest-energy states (in K units) of neutral  carbon, including the fine-structure levels and brightest NIR
and submillimeter lines discussed in the text. \mbox{Fine-structure} splittings are exaggerated for clarity.}
\label{Fig:energy_levels}
\end{figure}

\section{Spectroscopic parameters of neutral atomic carbon lines}
\label{app-spectroscopy}

Figure~\ref{Fig:energy_levels} shows a reduced energy-level diagram of neutral 
carbon including fine-structure splittings.
Table~\ref{Table_intensities} summarizes the spectroscopic parameters and line intensities of
the carbon lines discussed in the text (including PDR model predictions).


\begin{table*}[!h] 
\begin{center}
\caption{Spectroscopic parameters \citep[from NIST$^5$ and][]{Haris17} of the neutral carbon lines discussed in the text, intensities of the lines detected by JWST and ALMA in d203-506,
and model predictions.}  \label{Table_intensities}  
\small
\begin{tabular}{l l c c c c c c c @{\vrule height 10pt depth 5pt width 0pt}}    
\hline\hline       

 $\lambda^a$ ($\mu$m)  & Transition (u$-$l)                                                                                                     & $E_{\rm u}$/$k_{\rm B}$ (cm$^{-1}$) & $A_{\rm ul}$ (s$^{-1}$) &  $I_{\rm obs\,,\,\it{b}}^{\rm ON-OFF}$ (1$\sigma$ error) & $I^{\rm ON-OFF}_{\rm corr\,\it{b,\,c}}$ &
$I_{\rm PDR\,model,\,\it{b,\,d}}$  & $I_{\rm PDR\,model,\,\it{b,\,e}}$    \rule[-0.3cm]{0cm}{0.8cm}\ \\ \hline 
      609.1354         &  (2$p^2$)\,$^3$P$_1$\,--\,(2$p^2$)\,$^3$P$_0$       &           16.4 &    7.93E-08     &   1.62E-06 (9.77E-8)    & 1.62E-06   &  {(1.60--1.64)E-06} & {(1.60--1.64)E-06}\\             
      370.4151         &  (2$p^2$)\,$^3$P$_2$\,--\,(2$p^2$)\,$^3$P$_1$       &           43.4 &    2.65E-07     &                                          &            &  {(0.95--1.03)E-05}  &  {(0.95--1.03)E-05}\\\hline              
      0.982682         &  (2$p^2$)\,$^1$D$_2$\,--\,(2$p^2$)\,$^3$P$_1$       &        10192.6 &    7.30E-05     &                          &             &                &   \\
      0.985296         &  (2$p^2$)\,$^1$D$_2$\,--\,(2$p^2$)\,$^3$P$_2$       &        10192.6 &    2.20E-04     &                         &            &    &  \\\hline
      0.984 (sum)      &  (2$p^2$)\,$^1$D$_2$\,--\,(2$p^2$)\,$^3$P$_2$       &                &                 &   4.18E-04 (1.81E-05)   & 8.51E-04   &  {(2.14--4.90)E-04}  & {(1.07--2.04)E-05} \\ \hline
      0.872953         &  (2$p^2$)\,$^1$S$_0$--(2$p^2$)\,$^1$D$_2$           &        21648.0 &    6.00E-01     &       $\dagger$         &  $\dagger$ &  {(3.49--6.84)E-05}  & {(0.90--1.07)E-06}  &\\\hline

      1.068601         &  (3$p$)\,$^3$D$_2$\,--\,(3$s$)\,$^3$P$^{\rm o}_{1}$ &        69710.6 &   1.40E+07      &                         &            &  &  \\                    
      1.068827         &  (3$p$)\,$^3$D$_1$\,--\,(3$s$)\,$^3$P$^{\rm o}_{0}$ &        69744.1 &   1.04E+07      &                         &            &  & \\              
      1.069417         &  (3$p$)\,$^3$D$_3$\,--\,(3$s$)\,$^3$P$^{\rm o}_{2}$ &        69744.1 &   1.84E+07      &                         &            &  & \\              
      1.071027         &  (3$p$)\,$^3$D$_1$\,--\,(3$s$)\,$^3$P$^{\rm o}_{1}$ &        69689.5 &   7.50E+06      &                                     &            &   & \\              
      1.073247         &  (3$p$)\,$^3$D$_2$\,--\,(3$s$)\,$^3$P$^{\rm o}_{2}$ &        69710.6 &   4.40E+06      &                                     &            &   & \\              
      1.075692         &  (3$p$)\,$^3$D$_1$\,--\,(3$s$)\,$^3$P$^{\rm o}_{2}$ &        69689.5 &   4.80E+05      &                                     &            &   & \\ \hline  
      1.069 (sum)      &  (3$p$)\,$^3$D\,--\,$^3$P$^{0}$                        &                &                 &  1.67E-04 (3.67E-05)    &  3.23E-04  &  {(1.60--2.92)E-04} & {(2.36--2.51)E-06} \\ \hline
      1.175145         &  (3$d$)\,$^3$F$^{\rm o}_{2}$\,--\,(3$p$)\,$^3$D$_1$ &        78199.1 &   2.29E+07      &                                     &            &  & \\              
      1.175653         &  (3$d$)\,$^3$F$^{\rm o}_{4}$\,--\,(3$p$)\,$^3$D$_3$ &        78249.9 &   2.60E+07      &                                     &            &  & \\              
      1.175800         &  (3$d$)\,$^3$F$^{\rm o}_{3}$\,--\,(3$p$)\,$^3$D$_2$ &        78215.5 &   2.40E+07      &                                     &            &  & \\              
      1.178077         &  (3$d$)\,$^3$F$^{\rm o}_{2}$\,--\,(3$p$)\,$^3$D$_2$ &        78199.1 &   2.90E+06      &                                     &            &  & \\              
      1.180431         &  (3$d$)\,$^3$F$^{\rm o}_{3}$\,--\,(3$p$)\,$^3$D$_3$ &        78215.5 &   1.26E+06      &                                     &            &  & \\              
      1.182727         &  (3$d$)\,$^3$F$^{\rm o}_{2}$\,--\,(3$p$)\,$^3$D$_3$ &        78199.1 &   2.90E+03      &                                     &            &  & \\ \hline
      1.176 (sum)      &  (3$d$)\,$^3$F$^{\rm o}$\,--\,$^3$D                    &                          &       &   1.71E-05 (9.87E-06)   &  3.08E-05  &  {(3.61--5.96)E-05} & {(1.05--1.14)E-06} \\ \hline                    
      1.350596         &  (4$s$)\,$^3$P$^{\rm o}_{2}$\,--\,(3$p$)\,$^3$S$_1$ &        78148.1 &   6.60E+06      &                                             &                 &  & \\                   
      1.356337         &  (4$s$)\,$^3$P$^{\rm o}_{1}$\,--\,(3$p$)\,$^3$S$_1$ &        78116.7 &   6.50E+06      &                                             &             &  & \\                   
      1.358506         &  (4$s$)\,$^3$P$^{\rm o}_{0}$\,--\,(3$p$)\,$^3$S$_1$ &        78105.0 &   6.60E+06      &                                             &                 &  & \\\hline                   
      1.355 (sum)      &  (4$s$)\,$^3$P$^{\rm o}$\,--\,(3$p$)\,$^3$S         &                &                 &   2.82E-05 (5.75E-06)   & 4.55E-05   &  {(2.55--4.75)E-05} & {(1.71--1.60)E-07} \\ \hline                    
\hline
                                                                           
\end{tabular} 
\end{center} 
\normalsize
\tablefoot{$^a$Wavelengths in a vacuum. $\dagger$Observed with MUSE \citep{Haworth23}
{and  quoted also as the [\CI]\,8729\,\AA~line}.  $^b$In units of erg\,cm$^{-2}$\,s$^{-1}$\,sr$^{-1}$. 
$^c$Adopting $A_V$\,=\,1.5 mag of foreground extinction (see text).   $^d$Line intensity predicted by the reference PDR model (including C$^0$ FUV-pumping and C$^+$ recombination cascades) with
{$G_0$\,$\simeq$\,(1--2)$\times$10$^4$}, $n_{\rm H}$\,=\,10$^7$\,cm$^{-3}$, and $x_{\rm C}$\,=\,1.4$\times$10$^{-4}$.
$^e$ Same but neglecting FUV-pumping of C$^0$ lines.}                
\end{table*}      

\end{appendix}

\end{document}